\documentclass[a4paper,11pt]{article}
\usepackage{jheppub}

\usepackage{amsmath,amssymb}
\usepackage{amsthm}
\usepackage{ascmac}
\usepackage{graphicx}
\usepackage{subcaption}
\usepackage{physics}
\usepackage{comment}
\usepackage{esint}
\numberwithin{equation}{section}
\newcommand{\diag}{\operatorname{diag}}
\newcommand{\ii}{\mathrm{i}}
\newcommand{\ee}{\mathrm{e}}
\newcommand{\PO}{\mathrm{P}\,}
\newcommand{\SU}{\mathrm{SU}}
\newcommand{\U}{\mathrm{U}}
\newcommand{\pint}{-\!\!\!\!\!\!\int}

\newcommand{\Rep}[1]{\mathrm{Rep}\qty(#1)}

\newcommand{\tildet}{\tilde{t}}

\def\a{{\alpha}}
\def\b{{\beta}}
\def\d{{\delta}}

\def\l{{\lambda}}
\def\m{{\mu}}
\def\n{{\nu}}
\def\r{{\rho}}
\def\s{{\sigma}}

\def\th{{\theta}}

\AtBeginDocument{%
  \definecolor{red}{gray}{0.6}%
}

\newcommand{\shiro}{\raisebox{-2.5pt}{\scalebox{2.0}{$\circ$}}\;}
\newcommand{\kuro}{\raisebox{-2.5pt}{\scalebox{2.0}{$\bullet$}}\;}

\allowdisplaybreaks[4]

\title{Dominant Young Diagrams in Matrix Models and Partial Deconfinement}

\author[a]{Hidehiko Shimada,}
\emailAdd{shimada.hidehiko@gmail.com}

\author[b]{Hiromasa Watanabe}
\emailAdd{hiromasa.watanabe@keio.jp}

\affiliation[a]{National Institute of Technology, Akashi College,\\
679-3 Nishioka, Uozumi-cho, Akashi, Hyogo 674-8501, Japan}
\affiliation[b]{Department of Physics, and Research and Education Center for Natural Sciences, Keio University, \\
4-1-1 Hiyoshi, Yokohama, Kanagawa 223-8521, Japan}

\preprint{\today}

\abstract{
We discuss dominant representations (or Young diagrams),
in thermal matrix models with gauge symmetry
from the perspective of partial deconfinement.
We propose a prescription for defining the dominant representations for thermal matrix models with interaction terms.
As an explicit example, we consider the large-$N$ Gaussian matrix model. 
We obtain the 
Vershik--Kerov--Logan--Shepp (VKLS)
shape of the dominant Young diagrams through
a new analytic saddle-point analysis based on a mapping of the 
representation theory of U($\infty$) to free fermions in two spacetime dimensions.
This computation provides a direct derivation of the previously observed functional relation
between the shape of the dominant Young diagrams and the eigenvalue distribution
of the thermal holonomy: 
the position of the complex saddle point is naturally identified with the eigenvalue.
The dominant Young diagrams admit a natural interpretation in terms of partial deconfinement:
the number of rows in the dominant Young diagrams matches the size of the submatrix corresponding
to the deconfined subsector.
}

\begin{document}
\maketitle
\section{Introduction}
\label{sec:introduction}
Understanding the mechanism of confinement
and the thermal phase transition to the deconfined phase in gauge theories is an important theme~\cite{Wilson:1974sk,Polyakov:1978vu,Susskind:1979up}.
Apart from its significance in understanding the vacuum structure of Quantum Chromodynamics (QCD) and quark-gluon plasma under extreme conditions, 
the problem has attracted interest in large-$N$ gauge theories 
due to its relevance to 
black hole thermodynamics through the gauge/gravity duality~\cite{Maldacena:1997re,Aharony:1999ti}.

The large-$N$ limit plays the role of a thermodynamic limit; 
hence, phase transitions can take place even at finite spatial volumes.
An important example is the free limit of Yang-Mills theories with adjoint matter fields on a sphere~\cite{Sundborg:1999ue,Aharony:2003sx}.
For these models (and many others), there are three phases:
between the confined and deconfined phases,
there is an intermediate phase.
The confined and intermediate phases are separated by 
the Hagedorn transition~\cite{Hagedorn:1965st}, and the intermediate and deconfined phases by the GWW transition~\cite{Gross:1980he,Wadia:1980cp}.

In recent years, it has been proposed that the intermediate phase can be understood as a phase in which confined and deconfined degrees of freedom coexist~\cite{Hanada:2016pwv,Berenstein:2018lrm,Hanada:2018zxn}.
In this {\it partially deconfined} phase, 
these degrees of freedom coexist in the space of colour, not in the usual spacetime.
Thus, in a theory with $N \times N$ matrix degrees of freedom, $M \times M$ submatrix degrees of freedom are deconfined while the remaining degrees of freedom are confined. 

This interpretation succeeds in identifying an order parameter of the intermediate phase:
the size of the submatrix $M$ (or the ratio $M/N$).
The behaviour of various thermal observables, such as entropy and the Polyakov loop, as a function of 
$M$ can be quantitatively understood. 
Furthermore, the picture of partial deconfinement 
clarifies the crucial difference 
between the thermodynamic limits obtained via the large-volume limit and via the large-$N$ limit. 
In particular,
it explains how the peculiar thermodynamic property of the black hole --- the negativity of the specific heat --- can be realised in a large-$N$ gauge theory~\cite{Hanada:2016pwv}. 
The origin of this distinction lies in the all-to-all (``non-local'') 
nature of interactions in matrix models.
See Refs.~\cite{Hanada:2023rlk,Hanada:2022wcq} for reviews.

An essential feature of the picture of partial deconfinement 
is that the gauge invariance of the theory is fully maintained, 
despite the fact that it starts with identifying the deconfined sector 
as an $M \times M$ submatrix.
This is achieved as follows: 
for free theories, one begins with states in which only the submatrix degrees of freedom are excited 
(deconfined), and then applies the projection operator associated with the Gauss law constraint. 
After the projection, one is left with gauge-invariant states.
Thermodynamic quantities such as entropy are computed within this invariant Hilbert space,
yielding the correct result~\cite{Hanada:2018zxn}.
It was argued in Refs.~\cite{Watanabe:2020ufk,Hanada:2021ipb,Gautam:2022exf} 
that this picture naturally extends to interacting systems, in particular 
by employing colour-space analogues of wavepackets of one-particle states in interacting quantum 
field theories.

However, it is desirable to have a {\it manifestly} gauge-invariant 
characterisation of the partially deconfined phase, in particular, the identification
of the parameter $M$ which works seamlessly for interacting theories.
To this end, in this paper, we consider partial deconfinement from a
point of view based on the irreducible representations of the gauge group, $\SU(N)$. 
As is well known, an irreducible representation of $\SU(N)$ corresponds to a Young diagram whose number of rows is at most $N-1$.~\footnote{%
The correspondence between Young diagrams with at most $N-1$ rows and 
highest-weight representations of the Lie algebra $\mathrm{sl}(N, \mathbb{C})$ is explained, 
for example, in section 15.3 of \cite{FultonHarris}. 
The equivalence between these representations and the irreducible representations of
$\SU(N)$ follows from Weyl's well-known ``unitarian trick'' (see, for example, \cite{Varadarajan}, in particular section 4.11).}
By determining the Young diagrams which dominate the path integral, 
one can identify the parameter $M$ with the number of rows of the dominant Young diagrams.

The purpose of this paper is two-fold. 
The first is to discuss how one can introduce 
the idea of the dominant Young diagrams for general interacting theories.

The second is to clarify the relation between the dominant Young diagrams and 
another manifestly gauge-invariant order parameter which is well-defined for interacting theories: 
the eigenvalue distribution function $\rho(\theta)$ ($-\pi < \theta \le \pi$) of the Polyakov holonomy.
The system is in the (i) confined, (ii) partially deconfined, or (iii) deconfined phase when
$\r(\th)$ is (i) uniform, (ii) non-uniform but everywhere positive (ungapped), or (iii) 
non-uniform and vanishes over finite intervals in $(-\pi, \pi]$ (gapped), respectively.

It is natural to expect a direct relation between the shape
of the dominant Young diagrams and $\r(\th)$.
Indeed, Dutta and Gopakumar observed a simple functional relation between them by 
comparing the results of their computations~\cite{Dutta:2007ws}. 

In this paper, we give a thorough treatment of the standard example,
the Gaussian matrix model, focusing on the eigenvalue distribution,
dominant Young diagrams, and the relation between them.
We summarise what is known about the computation of $\r(\th)$ and the shape of the dominant Young diagrams,
as well as the relation between them,
and develop these results further.

We point out that the shape of the dominant Young diagrams in various phases 
can be understood naturally from the picture of partial deconfinement.
In particular, in the intermediate, partially deconfined regime, 
the shape is the so-called 
Vershik--Kerov--Logan--Shepp (VKLS)
shape~\cite{LS, VK} with $M$ rows,
consistent with the expectation that the number of rows of the dominant Young diagram
provides a gauge-invariant definition of $M$.

We then present a new analytic computation of $\r(\th)$ and the dominant Young diagrams.
This computation is based on a mapping ---developed by the Kyoto school of integrable systems 
and later applied to combinatorics by Okounkov and others 
(see, for example, \cite{Jimbo:1983if, Okounkov2001InfiniteWedge}) ---
from the representation theory of $\U(\infty)$ to free fermions in two spacetime dimensions.\footnote{%
A closely related application of this free-fermion framework can be found in Ref.~\cite{Murthy:2022ien}.%
}
Unlike the previous computation, no simplifying assumption on the couplings of the Gaussian matrix model is required.
Furthermore, the relation between $\r(\th)$ and the shape of the dominant Young diagram
arises directly from the saddle-point analysis; it is not necessary to compute the two observables
independently and compare them.

It is useful to recall that a crucial step in the large-$N$ saddle-point analysis
is the identification of an appropriate 
``large-$N$ collective field''~\cite{Jevicki:1979mb,Jevicki:1980zg}
--- a continuous function which represents the dominant configurations of the path integral in the large-$N$ limit.
The eigenvalue distribution $\rho(\theta)$ is a typical example of such a collective field~\cite{Brezin:1977sv, Aharony:2003sx}, 
while the continuous shape of dominant Young diagrams provides an alternative~\cite{Douglas:1993iia, Dutta:2007ws}.

These two collective fields are dual to each other.
Whereas the eigenvalues of a unitary matrix serve as coordinates on the group manifold $\SU(N)$,
the Young diagrams labelling irreducible representations correspond to momentum on $\SU(N)$, as is well known.
These coordinate and momentum descriptions are related to each other by the character expansion (the Peter-Weyl theorem), which is the Fourier transform on the group manifold.

In the saddle-point analysis of the dominant Young diagrams, one is forced to introduce another auxiliary
large-$N$ collective field $h(x)$ (to be defined in Sec.~\ref{sec:derivation1}),
as it appears in the explicit computation of the Boltzmann weight. 
Furthermore, it is $\tilde \r(h) = -\pdv{x}{h(x)}$, rather than the shape of the Young diagrams itself, that enters into the functional relation with $\r(\th)$.
We will see that our new computation gives a natural physical interpretation of these collective fields:
the continuous variable $h(x)$ corresponds to 
the fermion occupation level, 
whereas the shape of the Young diagrams is its bosonised counterpart.

We note that partial deconfinement from the viewpoint of dominant Young diagrams
was also studied by Berenstein and Yan~\cite{Berenstein:2023srv}.
For some models,
microcanonical ensembles --- rather than canonical ensembles ---
are suitable for describing the intermediate phase.
Motivated by this, they performed a combinatorial counting of gauge-invariant states of the Gaussian matrix model with a given energy, and found that the counting is dominated by Young diagrams with the VKLS shape~\cite{LS, VK}.

Another related recent work is that of O'Connor and Ramgoolam~\cite{OConnor:2026zlf}, who investigated the microcanonical thermodynamics of Gaussian matrix models.
Their analysis focuses on the regime in which the system has negative specific heat. 
To explain this feature, they analysed the ${\rm SO}(D)$ 
invariant sectors of the Gaussian $D$-matrix model.
Their setup differs from ours in that the ${\rm SO}(D)$ 
symmetry is gauged.  
Despite this difference, some of the representation-theoretic quantities discussed in their work are closely related to those considered in Secs.~\ref{sec:dominant_YD} and \ref{sec:GMM}.

Very recently, 
an interesting preprint~\cite{Ren:2026nrd} appeared on the arXiv,
in which the author discusses a mathematical dictionary between observables of various matrix models and spin chains. 
The tools and results used in that paper overlap with those in Sec.~\ref{sec:derivation2}.

We also note the recent independent work Ref.~\cite{deMelloKoch:2026jat}, which appeared on arXiv simultaneously with the first version of this paper and studies closely related aspects of partial deconfinement. 
While the authors of Ref.~\cite{deMelloKoch:2026jat} employ the canonical ensemble and investigate finite-$N$ corrections, the present work mainly focuses on the microcanonical treatment at large $N$. 
The two analyses provide different but complementary viewpoints on partial deconfinement in the representation basis.

This paper is organised as follows. 
We will discuss how to define the dominant representation for thermal matrix models in Sec.~\ref{sec:dominant_YD}.
Secs.~\ref{sec:GMM} and \ref{sec:derivation2} are devoted to the analysis of the Gaussian matrix model.
After introducing basic tools for the saddle-point analysis in Sec.~\ref{sec:GMM_as_UMM},
we review the direct saddle-point calculation of $\r(\th)$ and discuss the basic thermodynamic features
of the model in Sec.~\ref{sec:thermodynamics_GMM}.
In Sec.~\ref{sec:derivation1}, we summarise the computation
of the dominant Young diagrams via character expansion, and their relation to $\rho(\theta)$.
We then present, in Sec.~\ref{sec:derivation2}, a new saddle-point computation
of the VKLS shape of the dominant Young diagrams,
without the simplifying assumption used in the previous computation.    
Our approach gives a direct derivation of the relation between $\rho(\theta)$ 
and the shape of the dominant Young diagrams.
Finally, in Sec.~\ref{sec:summary}, we summarise our results and discuss further prospects.
Appendices summarise some basic facts we need about representations of $\SU(N)$, 
and present more detailed aspects in the computation of the dominant Young diagrams for the Gaussian matrix model.

\section{Defining dominant representations in thermal matrix models}
\label{sec:dominant_YD}

In this section, we consider a class of matrix models with the Lagrangian
\begin{equation}
    L = \sum_{I=1}^D \frac{1}{2}\tr(D_t X_I)^2 + V(X_I), 
    \qquad
    D_t X_I = \partial_t X_I - \ii [A_t, X_I] ,
\end{equation}
where $A_t$ is the temporal $\SU(N)$ gauge field, and 
$X_I$'s are hermitian $N\times N$ matrices (in the adjoint representation
of the gauge group).
The potential term $V(X_I)$ is arbitrary provided that it is invariant under
\begin{align}
    X_I \rightarrow U X_I U^{-1}
\end{align}
where $U$ is an $N\times N$ matrix with determinant unity.~\footnote{%
Note that the gauge symmetry is $\SU(N)$ rather than $\U(N)$, since the $\U(1)$ part does not
act on $X_I$.}

A crucial point is that we are considering a matrix model with gauge symmetry. 
The physical states contributing to the thermal partition function must satisfy the Gauss law constraint. 
This means that the states are invariant under the $\SU(N)$ gauge symmetry;
that is, they belong to the trivial representation.  

Nonetheless, it is known that in the large-$N$ limit,
thermal observables (in the deconfined and partially deconfined phases) are dominated by contributions from large Young diagrams
which can be approximated by a smooth profile~\cite{Douglas:1993iia, Dutta:2007ws, Berenstein:2018lrm}.

The purpose of this section is to discuss a prescription
for defining the smooth profile,
which we shall call the shape of dominant representations (of the gauge group $\SU(N)$)
or dominant Young diagrams, in a thermal ensemble
for a general theory with an arbitrary interaction term $V$.

Later in Sec.~\ref{sec:GMM}, we will focus on the Gaussian matrix model, 
$V(X_I)= \sum_I \frac{1}{2}\tr X_I^2 $, and show how the shape of dominant Young diagrams arises through explicit computations.

The thermal partition function of this model is
\begin{align}
    Z_\beta 
    =
    \int [\dd X \dd A_0]\ \exp\left(-\int_0^\beta \dd t\, L\right),
\end{align}
where $\beta$ is the inverse temperature, and the fields satisfy the standard periodic boundary condition, $X_I(\beta) = X_I(0)$ and $A_0(\beta) = A_0(0)$.
One can work in the gauge $A_0=0$. As is well known, a crucial role is played by the Polyakov holonomy $U \in \SU(N)$, defined as the path-ordered exponential
\begin{align}
    U= \PO \ee^{\ii\int_0^\beta A_0 \dd t}.
\end{align}
The existence of the Polyakov holonomy implies that, in the $A_0=0$ gauge, the $X_I$ fields satisfy the boundary condition $X_I(\beta) = U X_I(0) U^{-1}$.
One should integrate over $U$ to obtain $Z_\beta$,~\footnote{%
The Haar integral is normalised such that $\int \dd U = 1$.
}
\begin{align}
    Z_\beta = \int \dd U Z_\beta(U),
\end{align}
where $Z_\beta(U)$ is defined by the path integral with the boundary condition $X_I(\beta) = U X_I(0) U^{-1}$:
\begin{equation}
    Z_\beta(U) = \int_{X_I(\beta)=UX_I(0)U^{-1}} [\dd X] \ \exp\left(-\int_0^\beta \dd t\, L(X_I,\partial_t X_I)\right).
\end{equation}
Note that the Lagrangian $L$ in this expression does not contain $A_0$ (because of the $A_0=0$ gauge).
A crucial property of $Z_\beta(U)$ is its gauge invariance, or equivalently, that it is a class function:
\begin{align}
Z_\b(U) = Z_\b(VUV^{-1}),
\end{align}
where $V$ is an arbitrary element of $\SU(N)$.

The meaning of $Z_\beta(U)$ is as follows.
One of the most important observables in thermal Yang-Mills theory, 
the Polyakov loop (in the fundamental representation) \cite{Polyakov:1978vu}, can be expressed using $Z_\beta(U)$:
\begin{align}
    \left\langle \tr \PO \ee^{\ii\int_0^\beta A_0 \dd t} \right\rangle = \int \dd U \tr U Z_\beta (U).
\end{align}
Similarly, the multiply-wound Polyakov loop (with winding number $k$) is given by
\begin{align}
    \left\langle \tr \left(\PO \ee^{\ii\int_0^\beta A_0 \dd t} \right)^k \right\rangle = \int \dd U \tr U^k Z_\beta (U),
\end{align}
and the Polyakov loop in the representation $\lambda$ is given by 
\begin{align}
    \left\langle \tr R_\l\left(\PO \ee^{\ii\int_0^\beta A_0 \dd t} \right) \right\rangle 
    =
    \int \dd U \tr R_\lambda (U) Z_\beta (U)
    =
    \int \dd U s_\lambda(U) Z_\beta (U),
\end{align}
where $R_\lambda(U)$ is the representation matrix and $s_\lambda(U)=\tr R_\lambda(U)$ is the character associated with the representation $\lambda$.
As is well known, $s_\l(U)$ are given by the Schur polynomial of the eigenvalues of $U$ associated with $\l$.
(We have neglected an overall factor of $\frac1{Z_\beta}$ on the right-hand side.)
Thus, one can think of $Z_\beta(U)$ as a generating function for all Polyakov loops:
by expanding $Z_\beta(U)$ in terms of the characters of representations, 
we obtain the Polyakov loops corresponding to those representations.

These formulas can also be understood from the Hamiltonian formulation of the model.
We have
\begin{align}
Z_\b(U) = \Tr \hat U \ee^{-\b \hat H}.
\end{align}
Here, we denote by $\Tr$ the trace over the full Hilbert space of the theory (before imposing
the Gauss law constraint).
The insertion of $\hat U$, the representation of $U$ acting on the full Hilbert space,
takes care of the boundary condition $X_I(\b)= UX_I(0)U^{-1}$ in the path integral formulation.
(See also Ref.~\cite{Hanada:2020uvt}.)

The crucial observation is that the information contained in $Z_\beta(U)$ is insufficient to 
implement the idea of the dominant representation. 
The information in $Z_\beta(U)$ is in one-to-one correspondence with the ``end results'' 
of the computation of Polyakov loops in various representations. 
However, the dominant representations are, by definition, those which dominate in the intermediate stages of the computation of these observables. 
Thus, one cannot read off such internal information from $Z_\beta(U)$ alone, and there is no room for large representations to explicitly emerge from it.

In the following, we will show that one can read off 
the dominant representations, 
if we introduce a quantity, which we shall write $Z_\beta(U, V)$
which depends on {\it two} $N \times N$ unitary matrices, $U$ and $V$.
$Z_\b(U, V)$ is related to $Z(U)$ by
\begin{align}
    Z_\b(U, U) = Z_\b(U).
    \label{eq:ZUU_is_ZU}
\end{align}
As we shall see, 
$Z_\beta(U,V)$ serves as a better probe into the thermal property of the matrix model.

The following analogy may be useful.
We consider a wave in one compact spatial dimension $x$ described by a complex field $\phi(x)$, 
where the $x$-direction is analogous to $U$.
Let us say the total power is given by $\int \dd x\, \phi^*(x) \phi(x)$.
The local intensity $\phi^*(x)\phi(x)$
carries no information about which momentum modes are carrying the power.
If we know $\phi^*(x) \phi(y)$,
we can extract the power spectrum by carrying out Fourier expansions in $x$ and $y$ independently.
The quantities $Z_\beta(U)$ and $Z_\beta(U, V)$ are analogous to
$\phi^*(x)\phi(x)$ and $\phi^*(x) \phi(y)$, respectively.

In addition to the condition \eqref{eq:ZUU_is_ZU},
we require that $Z_\beta(U, V)$ is gauge invariant in both $U$ and $V$.
That is, for any $W \in \SU(N)$,
\begin{align}
    Z_\beta(W U W^{-1}, V) &= Z_\beta(U, V), 
    \label{eq:gauge_invariance_ZUV_L}
    \\
    Z_\beta(U, W V W^{-1}) &= Z_\beta(U, V).
    \label{eq:gauge_invariance_ZUV_R}
\end{align}
Then, one can perform the character expansion with respect to both $U$ and $V$ to obtain 
\begin{align}
    Z_\beta(U, V) = \sum_{\mu} \sum_{\nu} Z_{\mu \nu} s_\mu(U) s_\nu(V^{-1}),
\label{eq:partition_function_rep_generic}
\end{align}
where $s_\mu(U)$ denotes the character of $\SU(N)$ associated with the Young diagram $\mu$.
Here, we have adopted the convention of using $V^{-1}$ as the argument of $s_\nu$ rather than $V$.
Note that $s_\nu(V^{-1}) = s_\nu^*(V)$.

Clearly, the conditions \eqref{eq:ZUU_is_ZU}--\eqref{eq:gauge_invariance_ZUV_R} 
alone are not enough to fix $Z_\beta(U, V)$ uniquely.
It is an auxiliary quantity rather than a physical observable, and its construction
requires an additional prescription.

One can compute the thermal partition function in terms of $Z_\beta(U, V)$:
\begin{align}
    Z_\beta = \int \dd U\, Z_\beta(U, U) 
    = \int \dd U \sum_\mu \sum_\nu Z_{\mu \nu} s_\mu(U) s^*_\nu(U)
    = \sum_\mu Z_{\mu \mu},
\end{align}
where we have used the orthonormality of the characters, $\int \dd U\, s_\mu(U) s_\nu^*(U) = \delta_{\mu \nu}$.

It is natural to expect that, for a reasonably defined $Z_\beta(U, V)$, this
sum is dominated by large Young diagrams whose shapes are close to a specific continuous curve.
We shall refer to this shape as the shape of the dominant Young diagrams of the theory.
Note that there are exponentially many Young diagrams which are close to a given continuous curve.
One should determine the shape by minimising an effective action which 
incorporates both $Z_{\mu \mu}$ and this entropy effect.

One can also compute the Polyakov loop in a given representation $\rho$:
\begin{align}
    \int \dd U\, s_\rho(U) Z_\beta(U)
    &= \int \dd U\, s_\rho(U) Z_\beta(U, U) \nonumber \\
    &= \int \dd U \sum_\mu \sum_\nu Z_{\mu \nu} s_\rho(U) s_\mu(U) s^*_\nu(U) \nonumber \\
    &= \sum_\mu \sum_\nu c^\nu_{\rho \mu} Z_{\mu \nu}.
\label{eq:expectation_value_rep}
\end{align}
Here, $c^\nu_{\rho \mu}$ are the Littlewood-Richardson coefficients.
The coefficients count the multiplicity of the irreducible representation $\nu$
in the tensor product of the representations $\rho$ and $\mu$.
In terms of the characters, we have
$s_\rho(U) s_\mu(U) = \sum_\nu c^\nu_{\rho \mu} s_\nu(U)$.

By assumption, we consider a representation $\rho$ corresponding to a Young diagram 
which is much smaller than the dominant Young diagrams.
One can show that $c^\nu_{\rho \mu}$ is of order $1$ for given representations and 
is non-zero only if $\mu$ and $\nu$ have almost the same shape in this case.\footnote{%
See Sec. I.9 of \cite{Macdonald_sym_polynomial}.%
}
Hence, the shape of the dominant Young diagrams should be the same for the insertion of a small representation $\rho$ as for 
the thermal partition function.

A natural way to introduce $U$ and $V$ is 
to modify the adjoint action
of the gauge group $\SU(N)$ on the fields, $X_I \to U X_I U^{-1}$, to\footnote{%
A similar idea can be found in Refs.~\cite{Berenstein:2019esh,Berenstein:2023srv}.%
}
\begin{align}
    X_I \to U X_I V^{-1}.
\end{align}
The problem, of course, is that $U X_I V^{-1}$ is not hermitian in general. 
One can attempt to circumvent this issue in two ways:
(i) by analytic continuation, suitably rotating the integration contour in the complex plane, or
(ii) by doubling the degrees of freedom, replacing the hermitian matrix fields with complex matrix fields.

\subsection{Analytic continuation}

For the analytic continuation to work,
one needs to carefully examine the convergence of the integrals.
A full mathematical justification of the convergence is beyond the scope of this paper.
There may be several ways to implement the analytic continuation.
Below we present one of them based on the Hamiltonian formulation.

Let us recall that 
\begin{equation*}
    Z_\beta(U) = \Tr \left( \hat U \ee^{-\beta \hat{H}} \right),
\end{equation*}
where $\hat U$ is the operator representing how $U$ acts on the Hilbert space.
If we use the coordinate representation and denote the wavefunction by $\Psi(X_I)$,
we have
\begin{align}
\left(\hat U \Psi\right)(U X_I U^{-1}) = \Psi(X_I).
\end{align}
We now introduce two operators, $\hat U_{\rm L}$ and $\hat U_{\rm R}$, associated with an element $U \in \SU(N)$:
\begin{align}
\left(\hat{U}_{\rm L} \Psi\right)(U X_I) =& \Psi(X_I),
\\
\left(\hat{U}_{\rm R} \Psi\right)(X_I U^{-1}) =& \Psi(X_I), 
\end{align}
or equivalently,
\begin{align}
\left(\hat{U}_{\rm L} \Psi\right)(X_I) =& \Psi(U^{-1} X_I),
\\
\left(\hat{U}_{\rm R} \Psi\right)(X_I) =& \Psi(X_I U).
\end{align}
We have, by definition,
\begin{align}
\hat{U} = \hat{U}_{\rm L} \hat{U}_{\rm R}. \label{eq:hatU_hatUL_hatUR}
\end{align}
We also have
\begin{align}
\left(\hat{U'}_{\rm L} \hat{U}_{\rm L} \Psi\right) (X_I) 
=& \left(\hat{U}_{\rm L}\Psi\right)(U'^{-1} X_I)
= \Psi(U^{-1} U'^{-1} X_I)
= \Psi(\left(U' U\right)^{-1} X_I)
=\left(\widehat{U'U}_{\rm L} \Psi\right), \label{eq:group_property_L}
\\
\left(\hat{U'}_{\rm R} \hat{U}_{\rm R} \Psi\right)(X_I) 
=&\left(\hat{U}_{\rm R} \Psi\right)(X_I U') 
=\Psi(X_I U' U) 
=\left(\widehat{U'U}_{\rm R} \Psi\right)(X_I). \label{eq:group_property_R}
\end{align}

A crucial point is that, say, $U X_I$ is not hermitian. 
The wavefunction is originally defined as a function over hermitian matrices $X_I$,
which decays suitably when the components of $X_I$ go to $\pm \infty$.
In order to make sense of the definitions of $\hat{U}_{\rm L}$ and $\hat{U}_{\rm R}$,
one has to analytically continue the
wavefunction defined on the hermitian matrices.
This means, in particular, that the new wavefunction, say, $\hat{U}_{\rm L} \Psi$,
does not obey these decay properties.

Using these operators, one can define $Z_\b(U, V)$ by
\begin{equation}
    Z_\beta(U,V) = \Tr\left( \hat{U}_{\rm L} \hat{V}_{\rm R} \ee^{-\beta \hat{H}}\right).
    \label{eq:def_ZUV}
\end{equation}
The properties \eqref{eq:hatU_hatUL_hatUR}, \eqref{eq:group_property_L}, and 
\eqref{eq:group_property_R} assure the requirements for $Z_\b(U, V)$,  
\eqref{eq:ZUU_is_ZU},
\eqref{eq:gauge_invariance_ZUV_L}, and
\eqref{eq:gauge_invariance_ZUV_R}, respectively.
It is an important problem to study
whether the integral over the matrices $X_I$ converges in the trace operation of
\eqref{eq:def_ZUV}.

It seems that an analogous path-integral definition of $Z_\b(U, V)$ is not possible,
at least, not straightforwardly.
We can introduce two hermitian matrices, $A_0$ and $A_0'$, 
and use the covariant derivative $D_t X_I=  \partial_t X_I + \ii A_0 X_I - \ii X_I A'_0$ to write 
\begin{align}
    Z_\beta (U, V)
    = 
    \int [\dd X \dd A_0 \dd A'_0]\ \exp\left(-\int_0^\beta \dd t\ L\right),
\end{align}
where the periodic boundary conditions $X_I(\b)=X_I(0), A_0(\b)=A_0(0), A'_0(\b)=A'_0(0)$
as well as the constraints on the Polyakov holonomies, 
$U= \PO \ee^{\ii\int_0^\beta A_0 \dd t}$ and
$V= \PO \ee^{\ii\int_0^\beta A'_0 \dd t}$,
are understood.
Note that the action is not real.
While this definition yields $Z_\beta(U, V)$ satisfying \eqref{eq:ZUU_is_ZU},
it appears that \eqref{eq:gauge_invariance_ZUV_L} and \eqref{eq:gauge_invariance_ZUV_R} are not satisfied.
The reason is as follows.
The gauge invariance of $Z_\beta(U, V)$, expressed by
\eqref{eq:gauge_invariance_ZUV_L} and \eqref{eq:gauge_invariance_ZUV_R}, 
should follow from the gauge invariance of the path integral.
However, maintaining this gauge invariance is non-trivial.
Presumably, for a certain class of theories, one can redefine the action so that
it is gauge invariant under $X_I(t) \to U(t) X_I(t) V^{-1}(t)$.
However, the integration variables $X_I$ cease to be hermitian matrices after the gauge transformation,
requiring a $t$-dependent modification of the integration contour.
This breaks the gauge invariance of the path integral.
It is an interesting problem to find a path-integral definition of $Z_\beta(U, V)$.

\subsection{Doubling method}
\label{sec:doubling_method}
There is no convergence issue for 
the ``doubling'' method; however,
it works only for the free field theories. 

We introduce $D$ real hermitian matrix fields $Y_I$, and combine $X_I$ and $Y_I$ into complex matrix fields $\Phi_I$:
\begin{align}
    \Phi_I = \frac1{\sqrt2} \left(X_I + \ii Y_I\right).
\end{align}
For the $Y$-fields, we employ the same Lagrangian as that for the $X$-fields. 
Thus, the total action is 
\begin{align}
    S[\Phi_I] = \int_0^\b \dd t \left(L(X_I,\partial_t X_I) + L (Y_I, \partial_t Y_I)\right).
    \label{eq:action_doubled}
\end{align}
Note that we do not have gauge fields in this expression.

We now define $Z_\b(U, V)$ by
\begin{align}
    Z_\b(U, V) =& \left(\int_{\Phi_I(\b)= U \Phi_I(0) V^{-1}} [\dd X \dd Y]\ \ee^{-S[\Phi_I]}\right)^{\frac12}.
\end{align}
The path integral is defined with the boundary condition $\Phi_I(\b)= U \Phi_I(0) V^{-1}$.

We have taken the square root of the path integral on the right-hand side to compensate for the doubling of the degrees of freedom; consequently, for $U=V$, we recover 
\begin{align}
    Z_\b(U, U) = Z_\b(U).
\end{align}
The action is guaranteed to be real, and the path integral is real and positive definite. 
Hence, there is no ambiguity in defining the square root.

Since the kinetic and mass terms,
$\tr (\partial_t X_I)^2 + \tr (\partial_t Y_I)^2 = \tr (\partial_t \Phi_I^\dagger \partial_t \Phi_I)$
and $\tr X_I^2 + \tr Y_I^2 = \tr(\Phi_I^\dagger \Phi_I)$, are invariant under 
the $\SU(N)\times \SU(N)$ transformation, the free theory has the
$\SU(N)\times \SU(N)$ symmetry.
However, if there are interaction terms, the action \eqref{eq:action_doubled}
is not invariant under the transformation $\Phi_I \to U \Phi_I V^{-1}$. 
Hence, $Z_\b(U, V)$ defined by this method is not gauge invariant in general: 
the conditions \eqref{eq:gauge_invariance_ZUV_L} and \eqref{eq:gauge_invariance_ZUV_R} are not satisfied.

Equivalently, when $D$ is even, the $D$ hermitian matrices can be paired into $D_{\rm c}=D/2$ complex matrices
\begin{equation}
    \Phi_A=\frac{1}{\sqrt{2}}\qty(X_{2A-1}+\ii X_{2A}),
    \qquad
    A=1,\ldots,D_{\rm c}.
\end{equation}
For free theories we can define $Z(U, V)$ directly by
\begin{equation}
    Z_\beta(U,V)
    =
    \int_{\Phi_A(\beta) = U\Phi_A(0)V^{-1}} [\dd \Phi \dd \Phi^\dagger]\ \exp(-\int \dd t\ L[\Phi_A,\Phi_A^\dagger,\partial_t \Phi_A,\partial_t \Phi_A^\dagger]).
\end{equation}
See Appendix~\ref{sec:alg_approach_doubling} for an explicit computation of $Z_{\mu\nu}$ 
for the Gaussian matrix model based on this prescription.

\section{Analysis of the Gaussian matrix model via direct and character-expansion methods}
\label{sec:GMM}

In this and the next section, we give a thorough treatment of the Gaussian matrix model
in the large-$N$ limit, focusing on the dominant Young diagrams
and their relation to the eigenvalue distribution $\r(\th)$.
The analysis in this section summarises the standard direct and character-expansion approaches,
while in Sec.~\ref{sec:derivation2} we present new analytic derivation
of the shape of the dominant Young diagrams and its relation to $\r(\th)$.  

The thermal partition function defined by the path integral over gauge-invariant configurations
can be presented as a certain ordinary integral over unitary matrices~\cite{Sundborg:1999ue,Aharony:2003sx}.
This is reviewed in Sec.~\ref{sec:GMM_as_UMM}.

We then summarise the direct saddle-point computation of this unitary matrix integral 
using the eigenvalue density $\r(\th)$ in Sec.~\ref{sec:thermodynamics_GMM}.
We explain the three phases and how they are distinguished by the shape of $\r(\th)$~\cite{Sundborg:1999ue,Aharony:2003sx}.
We also briefly explain how these phases are understood in the picture of partial deconfinement
as the confined, partially deconfined, and deconfined phases~\cite{Hanada:2018zxn}.
The nature of the Hagedorn transition between the confined and partially deconfined phases
is also discussed in the language of state counting, following~\cite{Sundborg:1999ue,Aharony:2003sx}.

In Sec.~\ref{sec:derivation1}, following Dutta and Gopakumar~\cite{Dutta:2007ws}, 
we explain the evaluation of the unitary matrix integral as a sum over all irreducible 
representations of $\SU(N)$. 
The sum is dominated by large Young diagrams whose shapes are close to a specific continuous curve.
We summarise their computation of this curve, based on  
a simplified unitary matrix model in which certain coupling constants are set to zero.
We then explain the observation by Dutta and Gopakumar that
there is a simple functional relation between the dominant shape and $\r(\th)$.

\subsection{Hamiltonian formulation and the unitary matrix integral representation}
\label{sec:GMM_as_UMM}

We begin by explaining the basic properties of 
the Gaussian matrix model with gauge symmetry in the Hamiltonian formulation
(see also Refs.~\cite{Hanada:2019czd,Hanada:2023rlk}).
The Hamiltonian is given by 
\begin{equation}
    \hat{H}
    = 
    \frac{1}{2}\sum_{I=1}^D \Tr\qty(\hat{P}_I^2 + \hat{X}_I^2)
    =
    \sum_{I=1}^D \Tr \qty( \hat{A}^\dagger_I \hat{A}_I ) + \frac{DN^2}{2},
    \label{eq:H_GMM}
\end{equation}
where we have defined the creation and annihilation operators as
$\hat{A}_I^\dagger = \frac{\hat{X}_I - \ii \hat{P}_I}{\sqrt{2}}$
and 
$\hat{A}_I = \frac{\hat{X}_I + \ii \hat{P}_I}{\sqrt{2}}$.
Since the matrix operators $(\hat{P}_I)_{ij}$ and $(\hat{X}_I)_{ij}$ satisfy the canonical commutation relations $\comm{(\hat{X}_I)_{ij}}{(\hat{P}_J)_{kl}}=\ii\delta_{IJ}\delta_{il}\delta_{jk}$, they satisfy 
$\comm{(\hat{A}_I)_{ij}}{(\hat{A}_J^\dagger)_{kl}}=\delta_{IJ}\delta_{il}\delta_{jk}$.
Note that we have chosen the time scale such that $\hbar \omega=1$ for each oscillator.

The Hamiltonian is invariant under the gauge transformation $\hat U$, where $\hat X_I$ and $\hat P_I$ are transformed by a similarity transformation,
\begin{equation}
\begin{split}
    (\hat{P}_{I})_{ij} 
    \mapsto (U \hat{P}_{I} U^{-1})_{ij}
    =& \sum_{k,l} U_{ik} (\hat{P}_I)_{kl} U^{-1}_{lj},
    \\
    (\hat{X}_{I})_{ij} 
    \mapsto (U \hat{X}_{I} U^{-1})_{ij} 
    =& \sum_{k,l} U_{ik} (\hat{X}_I)_{kl} U^{-1}_{lj}.
\end{split}
\end{equation}
Here, $U$ is an arbitrary element of $\SU(N)$.
We have $\comm{\hat U}{\hat H}=0$.

As explained in Sec.~\ref{sec:dominant_YD},
the Gauss law (gauge-singlet) constraint must be imposed on the physical states.
The partition function is thus given by
\begin{equation*}
    Z_\beta
    =
    \int \dd{U} \Tr \qty(\hat U \ee^{-\beta \hat{H}}).
\end{equation*}
The integration over $U \in \SU(N)$ projects out non-singlet states.

We shall write $Z_\beta(U) = \Tr \qty(\hat U \ee^{-\beta \hat{H}})$ in terms of the eigenvalues of $U$.
Diagonalising $U$ by another unitary matrix $V \in \SU(N)$, we write
\begin{equation}
    U'= V U V^{-1} = \diag(\ee^{\ii \theta_1},\dots,\ee^{\ii\theta_N}).
    \label{eq:diagonalizing_U}
\end{equation}
We call $\th_i$ the eigenphases of $U$.
We have
\begin{align}
    Z_\beta(U) = \Tr \qty(\hat U \ee^{-\beta \hat H})
    = \Tr \qty(\hat V \hat U \ee^{-\beta \hat H}\hat V^{-1}) 
    = \Tr \qty(\hat U' \ee^{-\beta \hat H}).
\end{align}

In order to evaluate the trace, we note that
a basis of the Hilbert space is given by the Fock states,
\begin{align}
\prod_{I, i, j} 
\left((\hat{A^\dagger}_I)_{ij}\right)^{k_{I, i, j}}|0 \rangle,
\end{align}
where $k_{I, i, j}$ are non-negative integers (excitation numbers).
Using the action of the operator $\hat U'$ on $\hat{A}^\dagger_I$,
\begin{align}
\hat U' (\hat{A}^\dagger_I)_{ij} \hat U'{}^{-1} = \ee^{\ii(\theta_i - \theta_j)}(\hat{A}^\dagger_I)_{ij},
\end{align}
we have
\begin{align}
\hat U' 
\prod_{I, i, j} 
\left((\hat{A}^\dagger_I)_{ij}\right)^{k_{I, i, j}}|0 \rangle
=&
\prod_{I, i, j}\qty(\ee^{\ii (\th_i -\th_j)})^{k_{I, i, j} }
\left((\hat{A}^\dagger_I)_{ij}\right)^{k_{I, i, j}}|0 \rangle,
\\
\hat U' \ee^{-\b \hat H} \prod_{I, i, j} \left((\hat{A}^\dagger_I)_{ij}\right)^{k_{I, i, j}}|0 \rangle
=&
\prod_{I, i, j}\qty(\ee^{-\b+ \ii (\th_i -\th_j)})^{k_{I, i, j} }
\left((\hat{A}^\dagger_I)_{ij}\right)^{k_{I, i, j}}|0 \rangle,
\end{align}
where in the second line we have neglected the zero-point energies of the oscillators
since they only contribute an uninteresting overall factor to the partition function.
Thus, we obtain
\begin{align}
    Z_\beta(U) 
    =&
    \prod_{I,i, j} 
    \sum_{k_{I, i, j}=0}^\infty
    \qty(\ee^{-\b+ \ii (\th_i -\th_j)})^{k_{I, i, j} } 
    \\
    =&
    \prod_{i, j} 
    \qty(\sum_{k=0}^\infty
    \qty(\ee^{-\b+ \ii (\th_i -\th_j)})^{k})^D 
    \notag \\
    =&
    \qty(\prod_{i,j} \frac{1}{1-\ee^{-\beta+\ii(\theta_i-\theta_j)}})^D.
\end{align}

Introducing the effective potential of the matter sector, $V_{\rm mat} = - \log Z_\beta(U)$, we have
\begin{align}
    V_{\rm mat} 
    =
     \sum_{i,j} D \log \qty(1-\ee^{-\beta + \ii (\theta_i-\theta_j)})
    =
    -\sum_{i,j}\sum_{n=1}^\infty \frac{D}{n}\qty(\ee^{-\beta+\ii(\theta_i-\theta_j)})^n
    =
    -\sum_{n=1}^\infty \frac{D\ee^{-n\beta}}{n}\tr U^n \tr U^{-n},
\end{align}
where we have used
$
    \tr U^n = \sum_{i} \ee^{\ii n \theta_i}
$.
Thus, one can write the thermal partition function of the matrix model as an integral over a unitary matrix:
\begin{equation}
    Z_\beta = \int \dd U \exp\qty(\sum_{n=1}^\infty\frac{a_n(\beta)}{n}\tr U^n \tr U^{-n}).
    \label{eq:partition_function_generic}
\end{equation}
Here, we have introduced 
\begin{align}
a_n(\beta) = D\,\ee^{-n\beta}.
\label{eq:an_GMM}
\end{align}
For more general models (including Yang-Mills theory on $\mathrm{S}^3$ coupled to adjoint matter fields in the weak-coupling limit),
the same integral representation applies~\cite{Sundborg:1999ue,Aharony:2003sx}
provided that the coefficients $a_n(\beta)$ are suitably chosen.

One can replace the Haar integral by an integral over $\theta_i$ 
by introducing the Vandermonde determinant (up to an overall constant),
\begin{equation}
    \Delta(U) = \prod_{i < j} \abs{\ee^{\ii\theta_i}-\ee^{\ii\theta_j}}^2
     = \ee^{-\sum_{n=1}^\infty \frac{1}{n} \tr U^n \tr U^{-n}}.
\end{equation}

Hence, the thermal partition function finally takes the form 
\begin{equation}
    Z_\beta = \int \prod_i \dd \theta_i \exp\qty(-\sum_{n=1}^\infty\frac{1-a_n(\beta)}{n}\tr U^n \tr U^{-n}).
    \label{eq:Zbeta_an_with_Vandermonde}
\end{equation}
This will serve as the starting point for all saddle-point analyses presented below.

\subsection{
Direct saddle-point computation of the eigenvalue distribution
}
\label{sec:thermodynamics_GMM}

We now review the large-$N$ saddle-point computation of $\r(\th)$ based on the integral over the unitary 
matrix \eqref{eq:partition_function_generic} or \eqref{eq:Zbeta_an_with_Vandermonde},
and briefly overview the thermodynamics of the Gaussian matrix model.

We introduce the eigenphase distribution (called the Polyakov line phase distribution)
\begin{equation}
    \rho(\theta)
    =
    \frac{1}{N}\sum_{i=1}^N \delta(\theta-\theta_i),
    \qquad
    \int_{-\pi}^{\pi}\dd\theta\,\rho(\theta)=1 .
\end{equation}
The Fourier modes of $\rho(\theta)$ equal the normalised multiply-wound Polyakov loops,
\begin{equation}
    u_n
    =
    \frac{1}{N}\tr U^n 
    =
    \int_{-\pi}^{\pi}\dd\theta\,\rho(\theta)\ee^{\ii n\theta}
    .
    \label{eq:multiply-wound_PL}
\end{equation}
Conversely, one can write 
\begin{equation}
    \rho(\theta)
    =
    \frac{1}{2\pi}
    \qty[
        1+\sum_{n=1}^\infty
        \qty(u_n\ee^{-\ii n\theta}+u_n^*\ee^{\ii n\theta})
    ].
    \label{eq:fourier_expansion_of_rho}
\end{equation}
The reality of the distribution function $\rho(\theta)$ implies $u_{-n} = u_{n}^*$.
Furthermore, the model has the charge-conjugation symmetry $X_I \to X_I^*, A_0 \to -A_0$, which implies
that $\rho(\theta)$ is an even function, $u_{-n}=u_n$.

Treating $\rho(\theta)$ as the large-$N$ collective field, one can perform the saddle-point analysis~\cite{Sundborg:1999ue,Aharony:2003sx}. 
The free energy becomes
\begin{equation}
    \beta F = N^2\sum_{n=1}^\infty \frac{1 - a_n(\beta)}{n}\abs{u_n}^2,
    \label{eq:free_energy}
\end{equation}
up to a constant shift which does not depend on $u_n$.

The saddle-point configurations are given by $u_n$ which minimise $\b F$. 
One may treat $u_n$ as independent, except that there is a crucial constraint $\rho(\theta) \geq 0$.
Thus, there is a domain in the space of $u_n$ defined by this constraint and we are minimising 
the quadratic function $\b F$ within this domain.

For low temperatures, the coefficients of $\abs{u_n}^2$, $\frac{1-a_n(\beta)}{n}$, are all positive.
See \eqref{eq:an_GMM}
(In the low-temperature limit, $\beta \to \infty$,  $a_n(\beta) \to 0$.)
Hence, the saddle point is given by $u_n=0$ ($n=1, 2, \dots$).
This can be explained as the effect of the Vandermonde determinant:
it provides a repulsive force among eigenvalues, which dominates in this low-temperature regime.
As a result, the (completely) confined phase with a uniform Haar-random distribution 
$\rho(\theta) = 1/(2\pi)$ is realised as the saddle-point solution. 
In this phase, all nontrivial Polyakov loops vanish, and the free energy is of order $N^0$.

As the temperature increases, the coefficients $a_n(\beta)$ grow.
The first instability sets in at a temperature $\beta_{\mathrm{H}}$ defined by
\begin{equation}
    a_1(\beta_{\rm H})=1,
    \qquad
    \beta_{\rm H}=\log D,
\end{equation}
where the coefficient of $|u_1|^2$ vanishes.
Because of this, precisely at $\b = \b_{\rm H}$, 
the saddle-point configuration is characterised by a non-zero value of $u_1$, whereas
$u_2=u_3 = \dots = 0$.
Without loss of generality, we can choose $u_1 \ge 0$.
The eigenphase distribution is no longer constant and takes the form
\begin{equation}
    \rho(\theta)
    =
    \frac{1}{2\pi}
    \qty(1+2 u_1 \cos\theta).
    \label{eq:rho(theta)_partially_deconfined}
\end{equation}
The condition $\rho(\theta)\ge0$ implies that $u_1 \le \frac12$.
If $u_1 < \frac12$, $\r(\th) >0$ everywhere and the eigenphases are distributed all over $[-\pi, \pi]$.
For this reason, this non-uniform distribution $\r(\th)$ is called ungapped.

The transition from the phase with uniform $\rho(\theta)$ to the phase with non-uniform but ungapped $\rho(\theta)$ is 
identified with the Hagedorn transition~\cite{Hagedorn:1965st, Sundborg:1999ue, Aharony:2003sx}.
(See also the discussion at the end of this subsection regarding the relation to the 
Hagedorn behaviour of strings.)

The energy of the saddle point grows as $u_1$ increases from $0$ to $\frac12$.
These configurations labelled by $u_1$, or equivalently by the energy, constitute the intermediate phase found in Refs.~\cite{Sundborg:1999ue, Aharony:2003sx}. 
This phase was later interpreted as the partially deconfined phase in Refs.~\cite{Hanada:2016pwv,Berenstein:2018lrm,Hanada:2018zxn,Hanada:2023rlk}.
For the Gaussian matrix model, the configurations in this intermediate phase all have the same temperature, 
but different energies; hence, to treat them properly, the microcanonical rather than the canonical ensemble must be used.

If the temperature is further increased, we enter the third, completely deconfined phase.
Since $1-a_1(\b) < 0$, a larger value of $u_1$ is favoured to minimise the free energy.
However, in order to maintain $\rho(\th) \ge 0$, $u_2, u_3, u_4, \dots$ must become non-zero,
which in turn increases the free energy. 
The saddle-point configuration is determined from these two competing effects and
lies on the boundary of the domain in the space of $u_n$ defined by $\rho(\th) \geq 0$.
The distribution $\rho(\th)$ touches zero at some value of $\th$, 
and $\rho(\th)=0$ holds for a finite interval in this phase. 
For this reason, the distribution is called gapped.
The transition from the phase with non-uniform but ungapped $\r(\th)$ to the phase with gapped $\r(\th)$ is called 
the GWW transition~\cite{Gross:1980he,Wadia:1980cp}. 

For the Gaussian matrix model, 
the two phase transitions take place at the same temperature $T_{\rm GWW}=T_{\rm H}$,
and the microcanonical ensemble is suited to the treatment of 
the intermediate phase, parametrised by $u_1$ or the energy.
This is not a general feature of the intermediate phase;
for example, there are models in which $T_{\rm H}< T_{\rm GWW}$
and the intermediate phase can be described by the canonical ensemble~\cite{Aharony:2003sx,Schnitzer:2004qt}.

The interpretation of the intermediate phase as the partially deconfined phase
can be succinctly summarised by rewriting \eqref{eq:rho(theta)_partially_deconfined} as
\begin{equation}
    \rho(\theta)
    =
    \qty(1-\frac{M}{N})\cdot\frac{1}{2\pi}
    + 
    \frac{M}{N}\cdot\frac{1}{2\pi}(1+\cos\theta)
    =
    \qty(1-\frac{M}{N})\rho_{\rm con}(\theta)
    +
    \frac{M}{N}\rho_{\rm GWW}(\theta;M).
    \label{eq:rho_theta_partially_deconfined_interpreted}
\end{equation}
Here, $M$ denotes the size of the deconfined submatrix.
A crucial point is to identify the Polyakov loop $u_1$ and the ratio $M/N$ via 
\begin{align}
\frac{M}{N}= 2 u_1.
    \label{eq:MbyN_u1} 
\end{align}
The energy varies with the ratio $M/N$ as $E \propto \left(\frac{M}{N}\right)^2$, so that
any of the energy, the Polyakov loop $u_1$, or $M$
can be used as the order parameter for this intermediate phase.
The expression \eqref{eq:rho_theta_partially_deconfined_interpreted} signifies 
that we have two types of degrees of freedom---confined and deconfined---and $\r(\th)$ is given by their weighted average.
The distributions
$\rho_{\rm con}(\theta)=\frac1{2\pi}$
and $\rho_{\rm GWW}(\theta;M)=\frac1{2\pi}(1+\cos\th)$
are the eigenphase densities for  
the confined and the deconfined degrees of freedom, respectively.
The latter is denoted as such, since it coincides with the GWW point ($u_1=\frac12$) for the $\SU(M)$ theory.
See Refs.~\cite{Hanada:2023rlk,Hanada:2022wcq} for more complete reviews.

We conclude this subsection by explaining 
the understanding of the Hagedorn transition in the operator formalism~\cite{Sundborg:1999ue, Aharony:2003sx}.
Gauge-invariant states are created by trace operators built from the adjoint creation operators, such as
\begin{equation}
    \Tr\qty(\hat A_{I_1}^\dagger \hat A_{I_2}^\dagger \cdots \hat A_{I_L}^\dagger)\ket{0}.
    \label{eq:singletrace}
\end{equation}
Assuming $L \ll N$, the number of independent single-trace states of energy $E = L$ grows roughly as the number of cyclic words made from $D$ letters,
\begin{equation}
    \mathcal{N}(E) \sim \frac{D^L}{L}.
\end{equation}
Thus, the density of gauge-invariant states grows exponentially as
\begin{equation}
    \mathcal{N}(E) \sim \exp(\beta_{\rm H} E),
    \qquad
    \beta_{\rm H} = \log D .
\end{equation}
This exponential growth is the origin of the Hagedorn transition and gives the transition temperature agreeing with the condition $a_1(\beta_{\rm H})=D \ee^{-\beta_{\rm H}}=1$ found above.

\subsection{
Dominant Young diagrams via character expansion 
}
\label{sec:derivation1}

In this subsection, we first summarise the computation of the shape of the dominant Young diagrams,
following the work of  Dutta and Gopakumar~\cite{Dutta:2007ws}.
(The study of Young diagrams in the context of matrix models originated in \cite{Douglas:1993iia}.)
Here, we adopt the representation basis rather than the group-element basis used in the previous subsection.
We then point out that the resulting shape of the dominant Young diagram admits a natural
interpretation in the picture of partial deconfinement.

We begin with the thermal partition function \eqref{eq:partition_function_generic}
in the group-element (coordinate) basis, expressed as a Haar integral over $U$.
This can be rewritten in the representation (momentum) basis; 
one can carry out the Haar integral 
using various identities satisfied by the Schur polynomials~\cite{Dutta:2007ws}.
We obtain
\begin{equation}
    Z_\beta 
    =
    \sum_{\lambda} \sum_\mu \frac{\prod_{j=1}^\infty a_j^{\mu_j}}{z_\mu} \qty[\chi_\lambda(\tilde C_\mu)]^2.
    \label{eq:partition_function_chisq}
\end{equation}
The partition function is now written as a sum over irreducible representations $\lambda$ of $\SU(N)$,
where each $\lambda$ corresponds to a Young diagram with at most $N-1$ rows.
The summand involves quantities appearing in the representation theory of the permutation group.
Here, we have a sum over partitions $\mu$ of $K$, with $K=\sum_j j\mu_j$ equal to 
the number of boxes in $\lambda$.
We write
$
    z_\mu = \prod_{j=1}^\infty \mu_j! j^{\mu_j},
$
and denote by $\tilde C_\mu$
the conjugacy class of the permutation group $S_K$,
and by $\chi_\lambda(\tilde C_\mu)$ the character of the irreducible representation 
of $S_K$ corresponding to the Young diagram $\lambda$ evaluated at $\tilde C_\mu$.\footnote{%
Here we use the notation $\mu_j$ of Ref.~\cite{Dutta:2007ws}, which differs from our notation
elsewhere in our paper.%
}

Hereafter in this subsection, we focus on the low-temperature regime, $T < T_\mathrm{H}$ and $T = T_\mathrm{H}$.
They correspond to the confined phase and the partially deconfined phase, respectively,
as explained in Sec.~\ref{sec:thermodynamics_GMM}.
Accordingly, we now modify the model in order to facilitate the calculation.
Namely, we set $a_2 = a_3 = \cdots = 0$ and consider only the contribution from $a_1(\beta)$.
The rationale for this modification is as follows.

First, in these regimes, the Fourier components $u_n$ of $\r(\th)$ vanish for all $n \geq 2$.
The saddle-point analysis in Sec.~\ref{sec:thermodynamics_GMM} 
only depends on the condition $1-a_n > 0$ for $n \geq 2$
(see Eq.~\eqref{eq:free_energy}); 
setting $a_n = 0$ for $n \geq 2$ does not change the computation.

Second, this simplified model yields
certain profiles of dominant Young diagrams (the so-called VKLS shape).
The resulting shape turns out to be related to $\r(\th)$ via a simple functional relation.
This suggests that the simplified model captures correctly the physics of the full Gaussian matrix model.

However, we emphasise that there is no guarantee that the shape of the dominant Young diagram
is independent of parameters such as $a_2$. 
Incorporating these parameters into the analysis presented in this section remains an interesting open problem.\footnote{%
In Sec.~\ref{sec:derivation2}, we address this issue by a different approach:
we derive the VKLS shape without setting $a_2=a_3=\cdots=0$.%
}

Under this simplifying assumption $a_2 = a_3 = \cdots = 0$, 
the partition function \eqref{eq:partition_function_chisq} becomes
\begin{equation}
    Z(\beta) 
    =
    \sum_{m=0}^\infty \sum_{\lambda} \frac{1}{m!} \qty[d_\lambda(S_m)]^2 a_1^m \delta(\abs{\lambda}-m),
\end{equation}
where $m = \mu_1$. 
The delta function in this formula fixes $m$ to be the number of boxes of $\l$.
In this case, the character $\chi_\l$ 
equals the dimension of the representation $\l$ of the permutation group $S_m$,
which is given by the Frobenius-Weyl formula
\begin{equation}
    d_\lambda(S_m) = \frac{m!}{h_1!\cdots h_N!} \prod_{1 \le i < j \le N}(h_i - h_j),
\end{equation}
expressed in terms of the variables
\begin{equation}
    h_i = \lambda_i - i + N.
\end{equation}

The variables $h_i$ play a key role in the saddle-point analysis of this subsection.
Here, we are led to introduce these variables out of necessity
to express the Boltzmann weight via the representation
theory of the permutation group.
(Note that the variables $h_i$ are closely related to the hook lengths of the Young diagram, 
though their definition differs slightly from the standard one.)
We shall see in Sec.~\ref{sec:derivation2}
that these variables arise naturally when we analyse this model through
the mapping from the representation theory of $\mathrm{U}(\infty)$ 
to free fermions in two spacetime dimensions.
The strict inequalities
\begin{equation}
    h_1 > h_2 > \dots > h_N \ge 0
\end{equation}
reflect their underlying fermionic nature via the Pauli exclusion principle,
whereas the original variables $\lambda_i$ satisfying $\lambda_1 \ge \lambda_2 \ge \cdots \ge \lambda_N \ge 0$ constitute their bosonised counterpart.

We utilise the following variables (originally introduced in Ref.~\cite{Douglas:1993iia})
\begin{equation}
    x = \frac{i}{N} \in [0,1],
    \qquad
    \lambda(x) = \frac{\lambda_i}{N},
    \quad
    h(x) = \frac{h_i}{N} = \lambda(x) + 1 - x.
    \label{eq:def_x_lambda_h}
\end{equation}
In the large-$N$ limit, the partition function becomes
\begin{equation}
    Z(\beta) 
    =
    \int [\dd h(x)]\, \exp\qty(-N^2 S_{\rm eff}),
\end{equation}
where we have
\begin{align}
    S_{\rm eff} 
    = &
    - \pint \dd x \dd y \log\abs{h(x) - h(y)} 
    + 2 \int \dd x h(x)\log h(x)
    - \tilde m \log(a_1 \tilde m) - \tilde m - 1,
    \\
    \tilde m 
    =&\int_0^1\dd x\, \lambda(x)
    =\int_0^1\dd x\, h(x) - \frac{1}{2}, 
    \\
    m =& N^2 \tilde m. \label{eq:m_Nsquared_tilde_m}
\end{align}
This leads to the saddle-point equation 
\begin{equation}
    \fdv{S_{\rm eff}}{h} = 0 
    \quad
    \Rightarrow
    \quad
    \pint_0^1 \frac{\dd y}{h(x)-h(y)} = \log h(x) - \frac{1}{2}\log\qty(a_1 \tilde m).
    \label{eq:stationary_condition_h(x)}
\end{equation}

In order to solve this saddle-point equation systematically, it is useful to replace the monotonic function $h(x)$ by the density of the variables $h$. 
This is the standard continuum description used in, for example, random matrix theory and related Coulomb-gas problems: the saddle-point condition is rewritten as a singular integral equation in terms of a density function, which in turn is analysed by complex-analytic methods such as the resolvent~\cite{Brezin:1977sv}. 
Thus, we define
\begin{equation}
    \tilde \rho(h) = - \pdv{x(h)}{h}.
    \label{eq:tilde_rho_def}
\end{equation}
This density satisfies 
\begin{equation}
    \int_{h_{\min}}^{h_{\max}} \dd h\, \tilde \rho(h) = 1, 
    \qquad
    h_{\min} = h(x=1),
    \quad
    h_{\max} = h(x=0).
\end{equation}
Since $h_i-h_{i+1}=\lambda_i-\lambda_{i+1}+1\ge1$, we have
\begin{equation}
    \tilde \rho(h) \le 1.
    \label{eq:rho(h)_le_1}
\end{equation}
Using this density function, the stationary condition is rewritten as 
\begin{equation}
    \pint_{h_{\min}}^{h_{\max}} \frac{\dd h'\, \tilde \rho(h')}{h-h'}
    =
    \log\qty(\frac{h}{\xi}),
    \qquad
    \xi = \sqrt{a_1 \tilde m}.
    \label{eq:stationary_condition_rho(h)}
\end{equation}

The saddle-point equation \eqref{eq:stationary_condition_rho(h)} was 
solved in Ref.~\cite{Dutta:2007ws} by the usual resolvent method.
Depending on the value of $\xi$, there are (at least) two different classes of solutions.
The solution relevant for the confined and partially deconfined phases is
\begin{equation}
    \tilde\rho(h)
    =
    \begin{cases}
    \displaystyle
    1
    \quad&
    0 \le h < 1-2\xi,
    \\
    \displaystyle
    \frac{1}{\pi}\arccos(\frac{h-1}{2\xi}) 
    \quad&
    1-2\xi \le h \le 1+2\xi,
    \\
    \displaystyle
    0
    \quad&
    1+2\xi<h,
    \end{cases}
    \label{eq:rho(h)_PD_1}
\end{equation}
with 
\begin{equation}
    \tilde m = \xi^2
    \quad
    \Rightarrow
    \quad
    \xi = 0
    ~~
    \textrm{or}
    ~~
    a_1(\beta) = 1.
\end{equation}
The latter condition is satisfied at the critical temperature $T_{\rm c} = T_{\rm H} = T_{\rm GWW}$.
This solution is valid for $\xi \le \frac{1}{2}$; once $\xi$ exceeds this value, the constraint~\eqref{eq:rho(h)_le_1} triggers a Douglas--Kazakov-type phase transition~\cite{Douglas:1993iia} accompanied by a change of saddle.\footnote{%
After the transition, the system is in the (completely) deconfined phase.
The eigenphase distribution $\r(\th)$ is gapped. 
Correspondingly, the shape of the dominant Young diagrams has the following
characteristic: $\l(x)$ assumes a smooth profile for $0 \le x < 1$, where $\l(x)$ is a positive finite number.
Thus, the dominant Young diagrams are ``abruptly cut off'' around the $N$-th (more precisely, $N-1$-th) row;
the length of that row scales as $\l(1) N$.
See also the discussions in Secs.~\ref{sec:large-N-transition-comment} and \ref{sec:summary}.
}
The partially deconfined phase corresponds to $0 < \xi \le \frac12$.
The special degenerate case $\xi = 0$ corresponds to the completely confined phase where
the density $\tilde\rho(h)$ is a simple step function:
\begin{equation}
    \tilde\rho(h)
    =
    \begin{cases}
    \displaystyle
    1
    \quad&
    0 \le h < 1,
    \\
    \displaystyle
    0
    \quad&
    1 < h.
    \end{cases}
    \label{eq:rho(h)_C}
\end{equation}
The condition $\tilde m = \xi^2$, combined with \eqref{eq:m_Nsquared_tilde_m} means that the size (the total number of boxes) of the dominant Young diagrams in the large-$N$ limit scales as $N^2$.

In the completely confined case, we have $\xi=0$. 
In this case, we see that the partition function is dominated by the empty Young diagram, $\lambda = \emptyset$, \textit{i.e.}, the trivial representation of $\SU(N)$.

Finally, we compute the shape of the dominant Young diagrams from the saddle-point solution \eqref{eq:rho(h)_PD_1}.
It is convenient to introduce a new coordinate system:
\begin{equation}
    \lambda = \frac{\mathbf{X}+\mathbf{Y}}{\sqrt{2}},
    \qquad
    x = \frac{\mathbf{Y}-\mathbf{X}}{\sqrt{2}}.
\end{equation}
The density $\tilde \rho(h)$ becomes, from \eqref{eq:tilde_rho_def}, 
\begin{equation}
    \tilde \rho(h) = \frac{1}{2}\qty(1-\dv{\mathbf{Y}}{\mathbf{X}}).
\end{equation}
Integrating this relation, we obtain
\begin{align}
    \mathbf{Y}(\mathbf{X})
    &=
    \frac{2\xi}{\pi}
    +
    \int_0^{\mathbf{X}}\dd{\mathbf{X}'}
    \qty[1-2\tilde\rho(\sqrt{2}\mathbf{X}'+1)]
    \notag\\
    &=
    \begin{cases}
    \displaystyle
     \frac{2}{\pi}\qty(\sqrt{2\xi^2-\mathbf{X}^2}+\mathbf{X}\arcsin(\frac{\mathbf{X}}{\sqrt{2}\xi}))
     &
     \abs{\mathbf{X}} \le \sqrt{2}\xi,
     \\
     \displaystyle
     \abs{\mathbf{X}} 
     &
     \abs{\mathbf{X}} > \sqrt{2}\xi.
    \end{cases}
\end{align}
This is the so-called VKLS shape well-known in representation theory and combinatorics~\cite{LS, VK}.~\footnote{Its appearance in the Gross--Witten--Wadia model was discussed in Ref.~\cite{Chattopadhyay:2019pkl}.}
The area enclosed by this profile is 
\begin{equation}
    \int_{-\sqrt{2}\xi}^{\sqrt{2}\xi} \dd \mathbf{X} \,
    \qty(\mathbf{Y}(\mathbf{X}) - \abs{\mathbf{X}})
    =
    \xi^2
    =
    \tilde m.
\end{equation}
Since the area of each box in the $(\mathbf{X}, \mathbf{Y})$ coordinates is $1/N^2$,
we see that this area precisely recovers the total number of boxes in the dominant Young diagrams, $m=\tilde m N^2$.
See \eqref{eq:m_Nsquared_tilde_m}.

\paragraph{Relation between $\tilde\r(h)$ and $\r(\th)$}
%Remarkably, one finds that the dominant YD is implicitly but tightly connected to the eigenphase distribution in the group-element basis.
%The density functions in each basis hold a simple relation:
Comparing \eqref{eq:rho(h)_PD_1} with \eqref{eq:rho(theta)_partially_deconfined},
the authors of Ref.~\cite{Dutta:2007ws} made a remarkable observation 
that the two densities $\tilde\r(h)$ and $\r(\th)$---the collective fields in the representation basis and the group-element basis, respectively---obey the simple functional relation:
\begin{equation}
    \tilde\rho(h) = \frac{\theta}{\pi},
    \qquad
    \rho(\theta) = \frac{h}{2\pi}.
    \label{eq:Dutta-Gopakumar_rel}
\end{equation}
That is, the two densities are inverse functions of each other. 
Note that the following identification must be understood:
\begin{align}
\xi = u_1
\end{align}

This relation naturally suggests an underlying phase-space droplet description~\cite{Dutta:2007ws,Dutta:2015noa,Chattopadhyay:2017ckc,Chattopadhyay:2019pkl} in terms of an analogue of the Wigner function $w(\theta,h)$ satisfying\footnote{%
Note that this phase-space description is considered in colour space.
This is reminiscent of the wave-packet picture in colour space introduced 
from a different viewpoint in Ref.~\cite{Hanada:2021ipb}.}
\begin{align}
    \rho(\theta) &= \int \dd h\, w(\theta,h),
    \\
    \tilde\rho(h) &= \int \dd\theta\, w(\theta,h).
\end{align}
While this provides an appealing geometric intuition, 
the relation \eqref{eq:Dutta-Gopakumar_rel} was obtained only by comparing the two independent calculations, 
and its fundamental origin remains unclear.
In Sec.~\ref{sec:derivation2}, we present a step towards understanding the origin of this formula:
we give a direct saddle-point derivation of \eqref{eq:Dutta-Gopakumar_rel}, where the eigenphase $\th$ naturally emerges as the argument of the complex saddle point.

\paragraph{Interpretation of the shape of dominant Young diagrams from partial deconfinement}
The shape of the dominant Young diagrams naturally fits into the picture of partial deconfinement,
providing a description completely compatible with that in the group-element basis.
Crucially, the parameter $\xi$ specifying the shape coincides with the Polyakov loop, $\xi = u_1$.
Recall from Sec.~\ref{sec:thermodynamics_GMM} that $2 u_1 = M/N$ (see \eqref{eq:MbyN_u1}).
Thus, $\xi$ directly captures the fraction of the deconfined subsector.
More precisely, the number of rows of the dominant Young diagrams matches $M$.\footnote{%
This can be seen as follows.
One of the edges of the dominant Young diagrams corresponds 
to the point $(\mathbf{X}, \mathbf{Y}) = (-\sqrt{2}\xi, \sqrt{2}\xi)$, 
that is, $\l = 0$ and $x = 2\xi$. 
Using \eqref{eq:def_x_lambda_h}, this implies that the total number of rows of the Young diagrams equals 
$2 N \xi=2N u_1 = M$.% 
}
This confirms the expectation that the number of rows in the dominant Young diagrams provides a manifestly gauge-invariant definition of $M$.

While their size varies, the shape of the dominant Young diagrams maintains the VKLS shape throughout the partially deconfined phase.
Thus, the VKLS shape is the natural counterpart to the GWW eigenphase distribution $\rho_{\rm GWW}(\theta;M)$ in \eqref{eq:rho_theta_partially_deconfined_interpreted}.
These properties of the dominant Young diagrams suggest that they directly capture the deconfined degrees of freedom.~\footnote{%
A complementary canonical treatment of the partially deconfined regime was recently given in Ref.~\cite{deMelloKoch:2026jat} using the reduced model considered above. 
At the Hagedorn point, the canonical ensemble was shown to sample a continuum of configurations with different sizes of the deconfined subsector $M$, rather than selecting a single one.%
}

\section{
Free fermion approach to eigenvalue distribution and dominant Young diagrams
}
\label{sec:derivation2}
In this section, we present a computation of the dominant Young diagrams
and the eigenphase distribution $\r(\th)$ using a free-fermion description of representation theory of $\U(\infty)$~\cite{Jimbo:1983if, Okounkov2001InfiniteWedge}. 
The unitary matrix integral and the various associated quantities are represented as vacuum expectation values in the free-fermion Fock space.~\footnote{%
For a closely related approach to the unitary matrix integral, see \cite{Murthy:2022ien}.%
}
In this framework, the relation \eqref{eq:Dutta-Gopakumar_rel} 
between the representation basis and the group-element basis naturally emerges through a complex saddle-point
analysis: the argument of the complex saddle point directly matches the eigenphase $\th$.

In the large-$N$ limit, 
the saddle-point properties of the model \eqref{eq:partition_function_generic} can be deduced by 
studying another unitary matrix integral,\footnote{%
This integral is called ``the Circular Unitary Ensemble (CUE) with a potential deformation'' in 
the literature on random matrix theories.
In this section, we shall consider integrals over $\U(N)$ matrices rather than $\SU(N)$.
The distinction between them is irrelevant in the large-$N$ limit.%
}
\begin{equation}
    Z(t,\tildet)
    =
    \int_{\mathrm{U}(N)} \dd U 
    \exp\qty[
        \sum_{n=1}^\infty \qty(t_n \tr U^n + \tildet_n \tr U^{-n})
    ],
    \label{eq:CUE_pot-deformation}
\end{equation}
where $t = (t_1,t_2,\dots)$ and $\tildet = (\tildet_1, \tildet_2, \dots)$ are constant parameters
called the Miwa variables.
It is convenient to introduce matrices $X$ and $Y$ to parametrise the Miwa variables:
\begin{equation}
    t_n = \frac{1}{n} p_n(X) = \frac{1}{n}\tr X^n,
    \qquad
    \tildet_n = \frac{1}{n} p_n(Y) = \frac{1}{n}\tr Y^n.  
 \label{eq:t_X}
\end{equation}

We now elaborate on why it is possible to obtain observables of the original 
matrix model \eqref{eq:partition_function_generic} 
by computing corresponding ones of \eqref{eq:CUE_pot-deformation}.
The reason is that the large-$N$ saddle-point equation for $u_n = \frac{1}{N}\tr U^n$ associated with \eqref{eq:partition_function_generic} can be obtained self-consistently by solving the saddle-point equation associated with \eqref{eq:CUE_pot-deformation} and then imposing
\begin{align}
t_n = \tildet_n = \frac{a_n(\beta)u_n}{n}.
\label{eq:self_consistency_tn_vs_an_un}
\end{align}
See Sec.~5.3 of Ref.~\cite{Aharony:2003sx}.\footnote{%
The authors of Ref.~\cite{Aharony:2003sx} utilise this trick
and the result of Ref.~\cite{Jurkiewicz:1982iz}, in which the saddle-point solution is found 
for a general model of the type \eqref{eq:CUE_pot-deformation}, to obtain
the generic saddle-point configuration of $\r(\th)$, including that in the (completely) deconfined
phase.%
}

Actually, the relation between the unitary matrix models
\eqref{eq:partition_function_rep_generic} and \eqref{eq:CUE_pot-deformation} can be further
substantiated by the use of a Hubbard--Stratonovich transformation~\cite{Liu:2004vy}.
We shall show below first that the partition function of the original Gaussian matrix model
can be represented as an ``ensemble average''---a weighted average over Miwa variables with 
a specific weight---of the partition function \eqref{eq:CUE_pot-deformation}.
Then we show that the weighted average is self-averaging because 
it is dominated by the Miwa variables such that the condition
\eqref{eq:self_consistency_tn_vs_an_un}.

Consider the following Hubbard--Stratonovich transformation:
\begin{equation}
    \exp(\frac{a_n}{n}\abs{\tr U^n}^2)
    =
    \frac{n}{\pi a_n} \int \dd^2\Phi_n\,
    \exp(-\frac{n}{a_n}\abs{\Phi_n}^2+\Phi_n^* \tr U^n+\Phi_n \tr U^{-n}).
\end{equation}
Then the thermal partition function of the Gaussian matrix model can be rewritten as 
\begin{align}
    Z_{\rm GMM} 
    &=
    \int \dd U \exp\qty(\sum_{n=1}^\infty\frac{a_n(\beta)}{n}\tr U^n \tr U^{-n})
    \\
    &= 
    \int \dd U \prod_n \int \dd^2\Phi_n \frac{n}{\pi a_n} \exp(-\frac{n}{a_n}\abs{\Phi_n}^2+\Phi_n^* \tr U^n+\Phi_n \tr U^{-n})
    \\
    &= 
    \int \qty[\dd\Phi] \exp(-\sum_n \frac{n}{a_n}\abs{\Phi_n}^2) Z(\Phi,\Phi^*),
\end{align}
where $[\dd\Phi] = \prod_n \dd^2 \Phi_n \frac{n}{\pi a_n}$ and, 
$Z(\Phi,\Phi^*)$ are \eqref{eq:CUE_pot-deformation} evaluated at complex conjugate parameters
$\Phi_n$ and $\Phi_n^*$.\footnote{%
To avoid misunderstandings, we note that $\Phi_n$ are complex parameters,
and we are not using matrix representations of Miwa variables $X$, $Y$ here.}
Thus, we have
\begin{equation}
    Z_{\rm GMM}
    =
    \int \qty[\dd \Phi]\, \ee^{-S_{\rm HS}(\Phi)} Z(\Phi, \Phi^*),
\label{eq:ZGMM_as_ensemble_average}
\end{equation}
where
\begin{align}
S_{\rm HS}(\Phi) = \sum_n \frac{n}{a_n}\abs{\Phi_n}^2.
\label{eq:S_HS}
\end{align}
This means that the thermal partition function of the original matrix model
is given by a weighted average of the partition function \eqref{eq:CUE_pot-deformation}
with the weight \eqref{eq:S_HS}.

The Miwa variables $t_n, \tildet_n$ in \eqref{eq:CUE_pot-deformation} are external parameters to be fixed, 
whereas the auxiliary variables $\Phi_n$ introduced by the Hubbard--Stratonovich transformation are dummy variables to be integrated as in \eqref{eq:ZGMM_as_ensemble_average}. 
However, as we shall see presently, in the large-$N$ limit 
certain values of $\Phi_n$ have a dominant contribution to the weighted average \eqref{eq:ZGMM_as_ensemble_average}.

Writing $\Phi_n= N \tau_n$, we see that 
the standard 't Hooft counting associated with \eqref{eq:CUE_pot-deformation} implies that 
the $N\to \infty$ limit with fixed $\tau$ converges to a genus-zero free energy which is independent of $N$:
\begin{align}
\frac1{N^2} \log Z(\Phi, \Phi^*) = 
\frac1{N^2} \log Z(N \tau, N\tau^*) \to \mathcal{F}(\tau, \tau^*).
\end{align}
This yields 
\begin{equation}
    Z_{\rm GMM} 
    =
    \int \qty[\dd \tau]\, \ee^{N^2 S_{\rm eff}(\tau)},
\end{equation}
with 
\begin{equation}
    S_{\rm eff} 
    = 
    -\sum_{n} \frac{n}{a_n}\abs{\tau_n}^2 
    +
    \mathcal{F}(\tau,\tau^*).
\end{equation}
The factor of $N^2$ in the exponent implies that the weighted average \eqref{eq:ZGMM_as_ensemble_average}
in the large-$N$ limit
can be evaluated by \eqref{eq:CUE_pot-deformation} 
for the value of $\tau$ which minimises $S_{\rm eff}$.\footnote{%
Note that there is a flat direction associated with $\tau_1$ in the integration over the $\tau$ variables, 
at the temperature $T_{\text H}=T_{\text{GWW}}$. 
At this temperature, $\tau_1$ coincides with
the fundamental Polyakov loop, $u_1$.
This flat direction is precisely the origin of the order parameter of the partially deconfined phase,
and we have to employ the microcanonical ensemble in this phase,
as explained in Sec.~\ref{sec:thermodynamics_GMM}.%
}
Notice that 
\begin{equation}
    \pdv{\mathcal{F}}{\tau_n^*}
    = 
    \frac{1}{N^2} N\ev{\tr U^n} = u_n,
\end{equation}
where $\ev{\tr U^n}$ is the expectation value associated with \eqref{eq:CUE_pot-deformation}.
Hence, the stationary condition gives,
\begin{equation}
    0 
    = 
    \pdv{S_{\rm eff}}{\tau_n^*}
    =
    -\frac{n}{a_n}\tau_n + u_n 
    \qquad
    \Rightarrow
    \qquad
    \tau_n = \frac{a_n u_n}{n},
\end{equation}
reproducing the self-consistency condition \eqref{eq:self_consistency_tn_vs_an_un}.

These considerations allow us to conclude that the shape of the dominant Young diagrams
of the Gaussian matrix model coincides with the GWW model satisfying 
\eqref{eq:self_consistency_tn_vs_an_un}.

Now, we focus on the low-temperature regime $T \le T_{\rm H} = T_{\rm GWW}$.
In this regime, we already know, from the analysis of Sec.~\ref{sec:thermodynamics_GMM}, that 
$u_2=u_3=\dots=0$. 
Therefore, the self-consistency condition \eqref{eq:self_consistency_tn_vs_an_un} implies that
we can set $t_2=t_3=\dots=0$.\footnote{%
We emphasise that this is not a simplifying assumption.
Despite the similar appearance, this procedure is fundamentally different
from that in Sec.~\ref{sec:derivation1}, where we set $a_2=a_3=\dots=0$.
There, the simplified model was introduced as a tractable approximation to the original model.
Here, it is guaranteed that setting $t_2=t_3=\dots=0$ correctly yields the observables of the
original Gaussian matrix model with $a_2 \neq 0, a_3 \neq 0, \dots$.%
}
Consequently, the saddle-point analysis reduces to that of the GWW model~\cite{Gross:1980he,Wadia:1980cp},
\begin{equation}
    Z_{\rm GWW}(t) = \int_{\SU(N)} \dd U\, \ee^{V_{\rm GWW}[U]},
    \qquad
    V_{\rm GWW}[U] = t(\tr U + \tr U^\dagger),
    \label{eq:ZGWW}
\end{equation}
where we have written $t= t_1=\tildet_1$ with a slight abuse of the notation.\footnote{%
Note that the argument based on the Hubbard--Stratonovich transformation is also applicable
to the partially deconfined phase of more general theories,  
rather than just of the Gaussian matrix model, and implies that
the shape of the dominant Young diagrams is the VKLS shape.
The models include  
the free limit of Yang-Mills theories (with adjoint matter fields) on ${\mathrm S}^3$ 
considered in \cite{Sundborg:1999ue,Aharony:1999ti}.
This is presumably consistent with the well-known ``universality'' of the VKLS shape
in combinatorics and representation theory
(reminiscent of that of the Wigner's semi-circle law in random matrices).
See, \textit{e.g.},
the comment on the VKLS limit shape in \cite{BorodinOkounkovOlshanski2000} and references therein.%
}

Thus, our goal in this section is to obtain
the shape of the dominant Young diagrams of the GWW model (and hence that of the Gaussian matrix model)
and show that it is the VKLS shape.

Note that once the self-consistency condition \eqref{eq:self_consistency_tn_vs_an_un} is
established as the saddle-point condition of the Hubbard--Stratonovich
weighted average \eqref{eq:ZGMM_as_ensemble_average},
no separate self-consistency condition needs to be
verified for other observables;
the same saddle point $\tau$ governs the evaluation of them, provided
that the observable is not too large to shift the location of the dominant saddle.
It therefore suffices to solve \eqref{eq:self_consistency_tn_vs_an_un} once;
this is precisely the self-consistency condition of Ref.~\cite{Aharony:2003sx},
already solved there using $\rho(\theta)$: $u_{n\ge2}=0$, with $u_1$
given by the ungapped GWW solution, in the regime we are now studying, $T\le T_{\rm H}$.\footnote{%
As an independent check, Sec.~\ref{sec:algebraic_approach} reproduces the
same solution ($t=u_1, u_{n\ge2}=0$) directly from the free-fermion computation;
while reassuring, this check is not logically necessary for the conclusion
just reached.%
}

In Sec.~\ref{sec:boson/fermion_corresp} we summarise the mathematical 
machinery in which representation-theoretic constructs, such as the Schur polynomials $s_\l$,
are written in terms of the Fock states of free fermions in two spacetime dimensions.
In Sec.~\ref{sec:Schur_measure},
the unitary matrix model \eqref{eq:CUE_pot-deformation}, or equivalently,
the Schur measure, is cast into that language.
Sec.~\ref{sec:fermionic_computation_dominant_YD} presents the saddle-point computation
of $\tilde\r(h)$.
The direct connection to $\r(\th)$ emerges naturally in this computation, 
and the resulting $\tilde\r(h)$ gives the VKLS shape of the dominant Young diagrams.
Sec.~\ref{sec:algebraic_approach} deals with an alternative algebraic computation of $\r(\th)$ 
in the spirit of Sec.~\ref{sec:dominant_YD}.
Sec.~\ref{sec:large-N-transition-comment} contains comments on how the GWW transition 
may be treated in this framework, as well as its connection to the trace relations.

\subsection{Boson/Fermion correspondence}
\label{sec:boson/fermion_corresp}

\begin{figure}[t]
    \centering
    \includegraphics[width=0.65\linewidth]{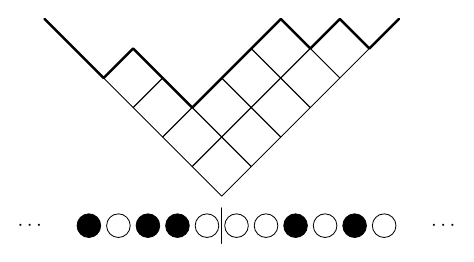}
    \caption{The Young diagram $\lambda = (5,4,1,1)$ and corresponding Maya diagram $\qty(\frac{9}{2},\frac{5}{2}|\frac{7}{2},\frac{1}{2})$. 
    }
    \label{fig:Maya-diagram}
\end{figure}

Our presentation here is based on the pioneering works~\cite{Jimbo:1983if,Okounkov2001InfiniteWedge}. 
Although these works originally considered ${\rm GL}(\infty)$ representations,
we use the language of $\U(N)$. 
Since we are interested in matrix models in the large-$N$ limit,
we maintain our notation and keep $N$ in some of the formulas in this section.
However, 
one should keep in mind that some of these formulas are invalid for large but finite $N$.
For example, a crucial difference between the strict large-$N$ limit and large but finite $N$ is that 
there is no restriction on the number of rows of Young diagrams in the former. 

We will first introduce {\it Maya diagrams}, which are in one-to-one correspondence with Young diagrams.
We then define the free-fermion theory in two spacetime dimensions, where 
each Maya diagram specifies a state in the Fock space.
We will see that the original Young diagrams naturally correspond to the bosonised counterpart 
of the free fermion.

We take the transpose of the Young diagram (exchanging the rows and columns) and rotate it by $45^\circ$ so that the upper-left corner comes to the bottom (see Fig.~\ref{fig:Maya-diagram}).
We then assign the left-ascending and right-ascending unit line segments (\textit{i.e.}, $\diagdown$ and $\diagup$) to $\kuro$ and $\shiro$, respectively.
As a result, we obtain an infinite sequence of black and white circles, 
which is the Maya diagram corresponding to the Young diagram.
We insert a vertical bar, $|$, to indicate the position of the corner.
Note that the left-most and right-most parts of a Maya diagram consist of infinite sequences of black and white circles, respectively.

We use a half-integer-valued label $r$ to specify the positions (which we call ``levels'' for reasons which will become clear in a moment) of the circles on the Maya diagram.
We can think of the vertical bar (or the corner) as being located at $r=0$.
The levels of the circles immediately to the right and left correspond to $r=\frac12$ and $r=-\frac12$, respectively.
Note that the sign assigned to a given level ($r > 0$ or $r < 0$) 
depends on the specific operator under consideration; see below for our precise conventions.

The Young diagram with zero boxes, associated with the trivial representation,
corresponds to the Maya diagram:
\begin{equation*}
    \cdots\kuro\kuro\kuro
    |~
    \shiro\shiro\shiro\cdots
    \ .
    \label{eq:Maya_diagram_empty}
\end{equation*}
We will later identify this Maya diagram with the Fock vacuum of the free fermion.

For any Maya diagram, by construction, the number of $\kuro$ to the right of the vertical bar is equal to the number of $\shiro$ to its left.
Let us denote this number by $d$. 
A Maya diagram can be specified by fixing the levels $\a_1 > \a_2 > \dots > \a_d > 0$ of 
$\kuro$ to the right, and $-\b_1 < - \b_2 < \dots < -\b_d < 0$ of $\shiro$ to the left of the vertical bar.
We use $(\a\mid\b) = (\a_1, \a_2, \dots, \a_d\mid\b_1, \b_2, \dots, \b_d)$ to denote the Maya diagram.
This notation is equivalent to the standard Frobenius representation (see, \textit{e.g.}, p.~3 of Ref.~\cite{Macdonald_sym_polynomial}) of the corresponding Young diagram, apart from the shift by $\frac12$ in the definition of the $\a$'s and $\b$'s.

We now identify the level $r$ as the mode number or spatial momentum of a free-fermion theory in two spacetime dimensions,
each Maya diagram representing a state in the Fock space of the theory.
We introduce the fermionic operators satisfying the Clifford algebra\footnote{%
Note that, in this notation, the adjoint operator of $\psi_r$ is $\psi^*_{-r}$.%
}
\begin{equation}
    \acomm{\psi_r}{\psi^*_s} = \delta_{r+s,0},
    \quad
    \acomm{\psi_r}{\psi_s}
    =
    \acomm{\psi^*_r}{\psi^*_s} 
    =
    0,
    \label{eq:anti-commutation}
\end{equation}
where $r, s \in \mathbb{Z}+\frac{1}{2}$.
Our convention is as follows:
the operator $\psi_r$ creates a ``particle'' ($\kuro$) at level $-r$,
and $\psi^*_s$ creates a ``hole'' ($\shiro$) at level $s$, respectively. 
Thus, pictorially, we have
\begin{align}
    \psi_r:& \textrm{  putting $\kuro$ at level $-r$},
    \\
    \psi^*_s:& \textrm{  putting $\shiro$ at level $s$}.
\end{align}
The relations $\psi_r^2 = (\psi^*_s)^2 = 0$ signify the Pauli exclusion principle.
The Fock vacuum is the state annihilated by the mode operators with positive levels:
\begin{equation}
    \psi_r\ket{\emptyset} = \psi^*_r\ket{\emptyset} = 0,
    \quad
    r > 0.
\end{equation}
Equivalently, $\psi_r$ ($r<0$) and $\psi^*_s$ ($s < 0$) are the creation operators of 
$\kuro$ and $\shiro$ at the levels $-r>0$ and $s<0$, respectively.
The Fock vacuum $\ket{\emptyset}$ is identified as
\begin{equation}
    \ket{\emptyset}
    \quad
    \Leftrightarrow
    \quad
    \cdots\kuro\kuro\kuro
    |~
    \shiro\shiro\shiro\cdots
    \ .
    \label{eq:Fock_vacuum_MD}
\end{equation}
The fermion levels to the left of the vertical bar are all occupied, while those to the right are empty.
Thus, the state can be understood as the Fermi (or Dirac) sea,
with the vertical bar indicating the Fermi surface separating the occupied and unoccupied levels.
Any state corresponding to the Young diagram $\lambda$ can be written as 
\begin{equation}
    \ket{\lambda} = \prod_{j=1}^d \psi^*_{-\beta_j}\psi_{-\alpha_j} \ket{\emptyset},
\end{equation}
with the Frobenius representation
$
    \lambda = \qty(\alpha_1,\dots,\alpha_d \mid \beta_1,\dots,\beta_d).
    %\label{eq:Frobenius_rep}
$
Thus, Maya diagrams provide a fermionic description of representations of $\U(N)$.\footnote{%
We consider the sector of the fermionic theory where total fermion number is zero.
For this sector, $\ket{\l}$ provides the complete set of basis.%
}
The trivial representation corresponds to the Fermi sea, and non-trivial representations are obtained by 
excitations around it.

The fermion fields corresponding to $\psi_r$ and $\psi^*_s$ can be
written in the notation often used in CFT as\footnote{%
See, for example, Sec.~10.2 of Ref.~\cite{Polchinski:1998rr}.
Note, however, that $\psi(z)$ here is a Dirac field, not subject to a Majorana condition.%
}
\begin{equation}
    \psi(z)=\sum_{r\in\mathbb{Z}+\frac{1}{2}}\psi_r z^{-r-\frac{1}{2}},
    \qquad
    \psi^*(z)=\sum_{r\in\mathbb{Z}+\frac{1}{2}}\psi^*_r z^{-r-\frac{1}{2}}.
    \label{eq:def_psi(z)}
\end{equation}
Restricting $z$ to the unit circle, $z=\ee^{\ii\theta}$, we can introduce the variable $\theta$, 
Fourier-conjugate to the level $r$, which plays an important role in the subsequent analysis.

We introduce the $\U(1)$ current $J(z)$ by
\begin{equation}
    J(z)=:\psi(z)\psi^*(z):
    =
    \sum_{n\in\mathbb{Z}}\rho_n z^{-n-1}.
\end{equation}
We obtain the bosonic operators $\rho_n$ by Fourier expanding $J(z)$ in terms of $\th$:
\begin{equation}
    J(\theta)=zJ(z)\big|_{z=\ee^{\ii\theta}}
    =
    \sum_{n\in\mathbb{Z}}\rho_n \ee^{-\ii n\theta},
    \qquad
    \rho_n
    =
    \int_{-\pi}^{\pi}\frac{\dd\theta}{2\pi}\,
    \ee^{\ii n\theta}J(\theta).
\end{equation}
In terms of the fermionic mode operators $\psi_r$ and $\psi^*_r$, we have
\begin{equation}
    \rho_n 
    = 
    \sum_{r\in\mathbb{Z}+\frac{1}{2}}
    \,:\psi_{-r}\psi^*_{r+n}: , 
\end{equation}
which satisfy the commutation relations of the Heisenberg algebra 
\begin{equation}
    \comm{\rho_n}{\rho_m} = n \delta_{n+m,0}.
    \label{eq:heisenberg_alg}
\end{equation}
These are standard definitions in the bosonisation prescription
(see, \textit{e.g.}, Ref.~\cite{Senechal:1999us} and Sec.~12.3 of Ref.~\cite{DiFrancesco:1997nk}).
Note that these algebraic relations take these simple forms only in the strict large-$N$ limit.

Positive modes, $\rho_{n}$ ($n>0$), annihilate the vacuum,
\begin{equation}
    \rho_n\ket{\emptyset}=0,
\end{equation}
A negative mode, $\rho_{-n}$ ($n>0$), excites particle--hole pairs, on the other hand;
the action of $\rho_{-n}$ moves a fermion in an occupied level in the Dirac sea to an empty level $n$ units to its right.  
The excitation of such a pair corresponds to a so-called hook representation, whose Young diagram 
consists of a single row and a single column sharing their first box.
In the Frobenius representation, this corresponds to $(\alpha\mid\beta)$ with $\alpha+\beta=n$.

It is convenient to define the vertex operators using $\rho_n$ as
\begin{equation}
    \Gamma_\pm(X) 
    = 
    \exp\left[\sum_{n=1}^\infty\frac{\rho_{\pm n}}{n}\tr X^n\right],
\end{equation}
where the matrix $X$ encodes the coupling constants $t_n$ via \eqref{eq:t_X}.
These operators create coherent states when acting on the Fock vacuum: for $n \in \mathbb{Z}_{>0}$,
\begin{gather}
    \ket{X} = \Gamma_{-}(X)\ket{\emptyset},
    \qquad
    \rho_{n} \ket{X} = \tr X^n \ket{X},
    \\
    \bra{X} = \bra{\emptyset}\Gamma_+(X),
    \qquad
    \bra{X}\rho_{-n} = \tr X^n \bra{X}.
\end{gather}
These properties follow from the commutation relations
\begin{align}
    \comm{\rho_n}{\Gamma_{-}(X)} = \tr X^n\, \Gamma_{-}(X),
    \qquad
    \comm{\rho_{-n}}{\Gamma_{+}(X)} = \tr X^n\, \Gamma_{+}(X),
    \label{eq:comm_vertex_bos-mode}
\end{align}
which can be derived using \eqref{eq:heisenberg_alg}.

Now, finally, we can establish the connection between the free-fermion theory and the representation theory of $\U(N)$:
the Schur polynomial $s_\lambda(X)$ associated with the representation $\lambda$ is given precisely by
\begin{equation}
    s_\lambda(X) = \braket{\lambda}{X} = \mel{\lambda}{\Gamma_{-}(X)}{\emptyset}.
    \label{eq:schur_pol_X}
\end{equation}
We also have
\begin{equation}
    s_\lambda(X) 
    = 
    \braket{X}{\lambda}
    =
    \mel{\emptyset}{\Gamma_+(X)}{\lambda},
\end{equation}
since $s_\l(X)$ is real-valued.

In the analysis below, we find it convenient to define 
a ``creation operator'' $\mathcal{A}^\dagger_\lambda$ 
which creates any state $\ket{\l}$ corresponding to a given Young diagram $\l$\footnote{%
To the best of our knowledge, the definition of this operator has not appeared in the literature.%
}:
\begin{equation}
    \ket{\lambda} = \mathcal{A}^\dagger_\lambda\ket{\emptyset}.
\end{equation}
The key input is the Frobenius formula \eqref{eq:Frobenius_formula}.
Using \eqref{eq:comm_vertex_bos-mode} and \eqref{eq:schur_pol_X}, the symmetric polynomials 
$s_\lambda(X)$ and $p_n(X)=\tr X^n$ have Fock-space representations, namely $s_\lambda(X) = \braket{X}{\lambda}$ and 
$p_n(X) = \tr X^n \braket{X}{\emptyset}=\mel{X}{\rho_{-n}}{\emptyset}$.
This leads to  
\begin{equation}
    \mathcal{A}^\dagger_\lambda 
    =
    \sum_{\substack{\mu \\ {\rm s.t.}~ \abs{\mu}= \abs{\lambda}}}
    \frac{\chi_\lambda(C_\mu)}{z_\mu} \rho_{-\mu},
    \quad
    \rho_{-\mu} 
    =
    \prod_{j=1}^{\ell(\mu)} \rho_{-\mu_j} . 
    \label{eq:creation_op_boson}
\end{equation}
Note that by acting with the adjoint operator on the coherent state, one obtains
\begin{equation}
    \mathcal{A}_\lambda\ket{X} 
    = 
    \sum_{\substack{\mu \\ {\rm s.t.}~ \abs{\mu}= \abs{\lambda}}}
    \frac{\chi_\lambda(C_\mu)}{z_\mu} \rho_{\mu}\ket{X}
    =
    s_\lambda(X) \ket{X}.
\end{equation}

This bosonic creation operator is particularly useful for computations involving 
the Littlewood--Richardson fusion coefficients of representations.
Since the left-hand side of \eqref{eq:fusion_schur} can be rewritten as
\begin{equation}
    s_\lambda(X) s_\mu(X) 
    =
    \mel{\mu}{\mathcal{A}_\lambda}{X}
    =
    \mel{X}{\mathcal{A}^\dagger_\lambda}{\mu},
\end{equation}
the fusion coefficients can be succinctly expressed as the matrix element of 
$\mathcal{A}^\dagger_\lambda$ between the states $\ket{\mu}$ and $\ket{\nu}$:
\begin{equation}
    c_{\lambda\mu}^\nu 
    %= \braket{\nu}{\lambda,\mu} 
    = \mel{\nu}{\mathcal{A}^\dagger_\lambda}{\mu}
    = \mel{\mu}{\mathcal{A}_\lambda}{\nu}
    .
    \label{eq:fusion_coeff_algebraic}
\end{equation}
Note that the important property of fusion coefficients, $c_{\lambda\mu}^\nu = c_{\lambda\bar{\nu}}^{\bar{\mu}}$, follows straightforwardly from this formula, where $\bar{\mu}$ and $\bar{\nu}$ denote the complex conjugate representations of $\mu$ and $\nu$, respectively.

\subsection{Schur measure and unitary matrix integral}
\label{sec:Schur_measure}
In the literature of combinatorics and representation theory, the unitary matrix
integral \eqref{eq:CUE_pot-deformation} is 
expressed in terms of the Schur measure~\cite{Okounkov2001InfiniteWedge},
a probability measure on the set of partitions.~\footnote{%
A related Schur-measure description and the large-representation limit were employed in Ref.~\cite{Betzios:2022pji} to study non-singlet matrix quantum mechanics and its relation to two-dimensional black-hole microstates.%
}
Here, we introduce this measure and establish its equivalence
to \eqref{eq:CUE_pot-deformation}.

The Schur measure~\cite{Okounkov2001InfiniteWedge} is defined by
\begin{equation}
    \mu_{\rm Schur}^{(X,Y)}(\lambda)
    %:= 
    = 
    \frac{1}{Z(X,Y)} s_\lambda(X) s_\lambda(Y),
    \label{eq:Schur_measure}
\end{equation}
where 
$X$ and $Y$ are matrices
corresponding to Miwa variables \eqref{eq:t_X}.
The expectation values of observables are defined by
\begin{equation}
    \ev{O(\lambda)}_{\rm Schur}^{(X,Y)} 
    =
    \frac{1}{Z(X,Y)} \sum_\lambda O(\lambda) s_\lambda(X) s_\lambda(Y)
    .
    \label{eq:ev_schur}
\end{equation}
The normalisation factor $Z(X,Y)$ is
\begin{equation}
    Z(X,Y) 
    = 
    \sum_\lambda s_\lambda(X) s_\lambda(Y) .
\end{equation}

We will use later that, by using the coherent states, one can write the partition function and
the expectation values as
\begin{align}
    Z(X,Y) =& \sum_\lambda s_\lambda(X) s_\lambda(Y) = \sum_\lambda \mel{\emptyset}{\Gamma_+(X)}{\lambda} \mel{\lambda}{\Gamma_-(Y)}{\emptyset} = \braket{X}{Y},
    \\
    \ev{O(\lambda)}_{\rm Schur}^{(X,Y)} 
    =&
   \frac{\sum_\lambda 
    \mel{\emptyset}{\Gamma_+(X)}{\lambda} O(\lambda) \mel{\lambda}{\Gamma_-(Y)}{\emptyset} }{Z(X,Y)}
    =
    \frac{\mel{X}{O}{Y}}{\braket{X}{Y}}
    .
\end{align}

The Frobenius formula \eqref{eq:Frobenius_formula} and 
the orthogonality condition for characters of the permutation group \eqref{eq:orthogonal_condition_character_perm}
lead to 
\begin{align}
    \sum_\lambda s_\lambda(X) s_\lambda(Y) \notag
    =
    \sum_\mu \frac{p_\mu(X) p_\mu(Y)}{z_\mu} \notag
    =
    \exp[\sum_{n=1}^\infty \frac{p_n(X) p_n(Y)}{n}]
    .
    \label{eq:schur_meas_cauchy_sum_formula}
\end{align}
This identity is known as the Cauchy sum formula.
Using this relation, 
the original partition function, initially expressed as an integral over $\U(N)$, can be rewritten as
\begin{align}
    Z(t,\tildet) 
    &= 
    \int_{\mathrm{U}(N)} \dd U
    \exp[
        \sum_{n=1}^\infty \frac{1}{n}\qty(
            p_n(X) p_n(U) + p_n(Y) p_n(U^{-1})
        )
    ]
    \\
    &=
    \sum_{\lambda,\mu}
    s_\lambda(X) s_\mu(Y) \int_{\mathrm{U}(N)} \dd U\, s_\lambda(U) s_\mu(U^{-1})
    \\
    &=
    \sum_\lambda s_\lambda(X) s_\lambda(Y) 
    = Z(X, Y).
\end{align}
In going from the second to the third line, we have used 
the orthonormality of the Schur polynomials \eqref{eq:orthogonal_condition_schur}.
This establishes the equivalence between the unitary matrix integral \eqref{eq:CUE_pot-deformation}
and the Schur measure.
This computation reflects the fact that the group-element basis (first line)
and the representation basis (third line) are Fourier-conjugate to each other.

We can also write the original partition function of the Gaussian matrix model in terms
of the Schur measure. 
The Hubbard--Stratonovich transformation \eqref{eq:ZGMM_as_ensemble_average} becomes,
\begin{equation}
    Z_{\rm GMM} 
    =
    \int \qty[\dd \Phi]\, \ee^{-S_{\rm HS}(\Phi)} \sum_{\lambda} s_\lambda(\Phi) s_\lambda(\Phi^*),
\end{equation}
or equivalently,
\begin{equation}
    Z_{\rm GMM}
    =
    \sum_\lambda W_\lambda,
    \qquad
    W_\lambda 
    =
    \int \qty[\dd \Phi]\, \ee^{-S_{\rm HS}(\Phi)} s_\lambda(\Phi) s_\lambda(\Phi^*).
\end{equation}

\subsection{Dominant Young diagrams}
\label{sec:fermionic_computation_dominant_YD}

Here, we perform a direct saddle-point calculation of the 
shape of the dominant Young diagrams.
The large-$N$ collective field we employ here is 
the density of the ``particles'' ($\kuro$) on the Maya diagram.
This density is equivalent to $\tilde\r(h)$ introduced in Sec.~\ref{sec:derivation1}.

One can show that a level $r$ is occupied by $\kuro$,
precisely when $r =\l_i -i +\frac12 $ for some $i=1, 2, \dots$.\footnote{%
This can be understood as follows. 
A particle ($\kuro$) corresponds to the rightmost box of row $i$ of $\l$,
located at $(\l_i, i)$ in the original coordinates in which the Young
diagram is drawn. Rotating the diagram by $45^\circ$, as in
Fig.~\ref{fig:Maya-diagram}, and taking the size of the box into account, we get
$r=\l_i-i+\frac12$, which is consistent with $r\in\mathbb Z+\frac12$.%
}
This expression for $r$ matches the variable $h_i=\lambda_i-i+N$ used in
Sec.~\ref{sec:derivation1}, up to a constant shift.
The precise identification of the two continuum densities
requires a discussion of the scaling limit; see below.

We can use the fermion number operator to probe 
whether there is a ``particle'' at level $r$:
\begin{equation}
    \psi_{-r}\psi^*_r \ket{\lambda}
    =
    \begin{cases}
        \ket{\lambda}
        &
        \text{for } \kuro \text{at level } r,
        \\
        0
        &
        \text{for } \shiro \text{at level } r.
    \end{cases} 
\end{equation}
Thus, we define the particle density in the Maya diagram as the expectation value \eqref{eq:ev_schur} of this occupation-number operator:
\begin{equation}
    \tilde\rho(r)
    =
    \frac{1}{Z(X,Y)}\mel{X}{\psi_{-r}\psi^*_r}{Y}.
    \label{eq:fermionic_density}
\end{equation}

The one-point function of the density operator \eqref{eq:fermionic_density} can be 
computed by using a special case of the already known determinant formula~\cite{Okounkov2001InfiniteWedge}
\begin{equation}
    \tilde\rho(r_1,r_2,\dots,r_n) = \det_{1\le i,j\le n}\qty[K(r_i,r_j)]
\end{equation}
for the $n$-point density correlation function in the momentum space.
Namely, we have
\begin{align}
\tilde\rho (r) = K(r, r).
\end{align}
Here, the kernel is given by the contour integral~\footnote{%
Note that the integrand is a holomorphic function both in the $z$- and $w$- planes except at $z=0$ and $w=0$
because $r, s \in \mathbb Z+\frac12$.%
}
\begin{equation}
    K(r,s)
    =
    \oiint \frac{{\rm d}z{\rm d}w}{(2\pi\ii)^2} \frac{{\rm e}^{V_+(z;t)-V_-(z;\tilde t)-V_+(w;t)+V_-(w;\tilde t)}}{z-w}\frac{1}{z^{r+1/2}w^{-s+1/2}},
    \label{eq:kernel_integral}
\end{equation}
where
\begin{equation}
    V_\pm(z; t) 
    = 
    \sum_{n=1}^\infty t_n z^{\pm n}.
\end{equation}
The contours for the $z$- and $w$-integrations are circles of radius $1+\epsilon$ and $1-\epsilon$ ($\epsilon > 0$), respectively, both oriented counterclockwise.

We now specialise to the GWW model by setting
\begin{equation}
    t_1 = \tilde{t}_1 = L t,
    \qquad
    t_n = \tilde{t}_n = 0 \quad (n \ge 2).
\end{equation}
Following the formulation in Refs.~\cite{walsh:tel-03969540,BeteaBouttierWalsh:2024},
we have introduced 
a large parameter $L$ which controls the size of the Young diagrams and
allows us to define the scaling limit. 
This parameter is needed because we are working in the strict large-$N$ limit in this section.
The parameter $L$ plays the role played by the matrix size $N$ in Secs.~\ref{sec:thermodynamics_GMM} and \ref{sec:derivation1}.
To take the continuum limit, we scale the level $r$ as
\begin{equation}
    r = L h + O(1).
\end{equation}
We have, in this limit,
\begin{equation}
    V_+(z;t) - V_-(z;\tilde{t})
    =
    L\, t \left( z - z^{-1} \right).
\end{equation}
The continuous variable $h$ is the same variable $h$ introduced in Sec.~\ref{sec:derivation1}, up to an overall shift.

To compute the one-point function $\tilde\rho(r)=K(r,r)$, we first evaluate a
point-split form, $K(r_1,r_2)$, by keeping the two levels separated by an $O(1)$ quantity 
$\delta\in\mathbb{Z}$:
\begin{equation}
    r_1 = \lfloor Lh\rfloor + \frac{1}{2},
    \qquad
    r_2 = r_1 + \delta .
\end{equation}
Then $r_1/L,r_2/L\to h$, but $r_2-r_1$ is kept finite. 
The diagonal kernel $\tilde\r(r)$ is recovered at the end by taking
$\delta\to0$.

We now evaluate $K(r_1, r_2)$ by computing the integral \eqref{eq:kernel_integral} 
in the large-$L$ limit~\cite{okounkov_correlation_2003,
okounkov_symmetric_2003,BeteaBouttierWalsh:2024}.
We first locate the saddle points of the integrations over $z$ and $w$ in \eqref{eq:kernel_integral}.
Let us write 
the $z$-dependent part of the exponent in \eqref{eq:kernel_integral} as $L S_{\rm GWW}$.
In the scaling limit, it reduces to
\begin{equation}
    S_{\rm GWW}(z;h)
    =
    t\qty(z-z^{-1})-h\log z.
\end{equation}
We can find the saddle points by the stationarity condition,
\begin{equation}
    \partial_z S_{\rm GWW}(z;h)=0
    \qquad
    \Rightarrow
    \qquad
    t z^2 - h z + t = 0 ,
\end{equation}
which yields
\begin{equation}
    z_\pm(h)
    =
    \frac{h\pm\sqrt{h^2-4t^2}}{2t}.
    \label{eq:saddle_points_kernel}
\end{equation}
In the large-$L$ limit, the $w$-dependent part takes the same form, apart from an overall factor of $-1$.
Hence, the saddle points for the $w$-integration are also given by \eqref{eq:saddle_points_kernel}.
Due to this factor of $-1$, 
the steepest descent path of the $w$-integration passing through each saddle point 
intersects that of the $z$-integration orthogonally at the saddle point.

\begin{figure}[t]
    \centering
    \includegraphics{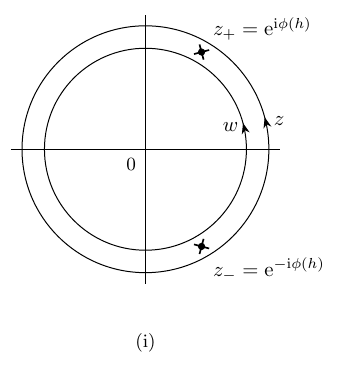}%
    \includegraphics{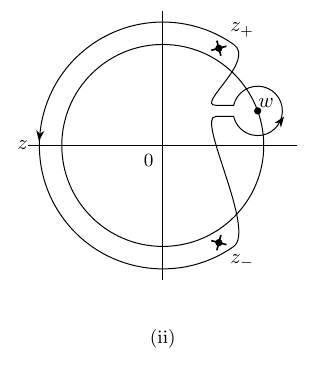}%
    \includegraphics{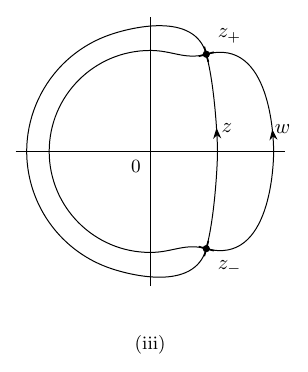}
    \caption{
    The saddle points $z_\pm$ lie on the unit circle. 
    Steepest descent directions are indicated by crosses.
    (i) The original integration contours for $z$ and $w$.
    (ii) The integration contour for $z$ crossing the position of $w$, when $w$ lies on the arc between $z_+$ and $z_-$. 
    (iii) The final ``steepest descent contours'' for the $z$- and $w$- integrations.
    \label{fig:contours}
    }
\end{figure}

We will first study the region $\abs{h}<2t$, in which the two saddle points are complex conjugate to each other and lie on the unit circle:
\begin{equation}
    z_\pm(h)
    =
    \frac{h}{2t}
    \pm
    \ii\sqrt{1-\frac{h^2}{4t^2}}
    =
    \ee^{\pm\ii\phi(h)} ,
    \qquad
    \phi(h)=\arccos\qty(\frac{h}{2t}).
    \label{eq:saddle-pt_phi}
\end{equation}
Note that $0 < \phi(h) <\pi$.
By evaluating $\partial_z^2 S_{\mathrm{GWW}}(z)$, one can show that
the steepest descent direction of the $z$-integration at $z_+$ ($z_-$)
is parallel to a line which intersects with the real axis at an angle of
$\phi(h)+\frac\pi4$ $\qty(-\phi(h)+\frac\pi4)$.
See Fig.~\ref{fig:contours}.

We can evaluate
the integral~\eqref{eq:kernel_integral} in the limit $L\to\infty$
by deforming the integration contours over $z$ and $w$ to the ``steepest-descent contours''
shown in Fig.~\ref{fig:contours}~(iii)\footnote{%
Strictly speaking, the true steepest descent contours are constructed from curves along which $\Im S$ is constant, with $\Re S$ increasing towards a saddle point along one branch and decreasing away from it along another.
In the present case, each contour consists of two such curves connecting $z=0$ and $z=\infty$ (or $w=0$ and $w=\infty$).
The contours in Fig.~\ref{fig:contours}~(iii) are instead required only to (a) coincide with the true steepest descent direction locally at each saddle point, so that the evaluation of the contribution from the saddle point is straightforward, and (b) be deformable into the true steepest descent contour without crossing any singularities and hence give the same integral.
It is more convenient to use contours satisfying only (a) and (b) when discussing the deformation of contours.%
}.
A crucial point is that the double integral over these contours is suppressed
in the $L\to\infty$ limit.
The suppression is power-law ($\sim \frac1L$) rather than exponential,
since $\Re S_{\mathrm{GWW}}=0$ at the saddle points.
Each of the $z$- and $w$- integrations produces a
factor of $\frac1{\sqrt L}$.
(For a similar computation, see, \textit{e.g.} (3.37) of Ref.~\cite{BeteaBouttierWalsh:2024}.)
There is, however, a contribution from the residue at the pole $z=w$,
since the two contours must cross each other during the deformation.
To be more precise, consider deforming the $z$-integration contour, say, for a fixed value of $w$,
before deforming the $w$-integration contour.
If $w$ lies on the arc $\mathcal{C}_h$ of the unit circle connecting the two saddle points (passing through $z=1$),
$
    \mathcal C_h=\{\,z=\ee^{\ii\varphi}\mid -\phi(h)\le\varphi\le\phi(h)\,\},
$
the contour for the $z$-integration necessarily 
crosses the pole at $z=w$, as illustrated in Fig.~\ref{fig:contours}~(ii).

In this contribution from the residue, the exponential factors in \eqref{eq:kernel_integral} cancel.  
Thus, in the scaling limit described above, the kernel is reduced to the one-dimensional integral
\begin{equation}
    K(r_1,r_2)
    \sim
    \frac{1}{2\pi \ii}
    \int_{\mathcal C_h} w^{r_2-r_1-1}\dd w .
    \label{eq:kernel_contribution_from_pole}
\end{equation}
Computing the integral with this parametrisation, we obtain
\begin{equation}
    K(r_1,r_1+\delta)
    \sim
    \int_{-\phi(h)}^{\phi(h)}
    \frac{\dd\varphi}{2\pi}\, \ee^{\ii\varphi\delta}
    =
    \frac{\sin\qty(\phi(h)\delta)}{\pi\delta}
    \qquad
    (\delta\ne0).
\end{equation}
Note that the contribution from the two conjugate saddle points
$\ee^{\ii\phi(h)\d}-\ee^{-\ii\phi(h)\d}$ is the origin of the familiar 
sine-kernel.

Taking the limit $\delta \to 0$, we finally obtain the continuum particle density $\tilde\r(h)$ on the Maya diagram:
\begin{equation}
    \tilde\rho(h)
    =
    \lim_{L\to\infty}K(\lfloor Lh\rfloor,\lfloor Lh\rfloor)
    =
    \frac{\phi(h)}{\pi}.
    \label{eq:tilde_rho_saddle-pt}
\end{equation}
Thus, the density is determined by the argument of the complex saddle point.

This is our key result.
As discussed around \eqref{eq:def_psi(z)}, the angular variable $\th = \arg z$ is Fourier-conjugate to the fermion level $r$ on the Maya diagram.
For the saddle point \eqref{eq:saddle-pt_phi}, this variable is directly given by $\th=\phi(h)$.
The angle $\th$ dual to $r$ should be identified with the eigenphase of $U$.
Roughly speaking, this is because the level $r$---or equivalently, $h$---labels the irreducible Young diagrams, which are in turn dual to the coordinates on the group manifold.
More precisely, 
this identification is strongly supported by the following argument.
Recall that the Schur polynomial possesses the determinant form
\begin{equation}
	s_\lambda(U)
    	=
	\frac{\underset{1\le i,j \le N}{\det}(\ee^{\ii(\lambda_j-j+N)\th_i})}{\underset{1\le i<j \le N}{\prod}(\ee^{\ii\theta_i} - \ee^{\ii\theta_j})},
\end{equation}
where the eigenphases $\th_i$ are paired with the fermionic variables $h_j = \lambda_j - j + N$.
Schematically, the ``wavefunction'' takes the form $\ee^{\ii h_i \th_i}$. 
This supports our identification of the variable $\th$ in \eqref{eq:tilde_rho_saddle-pt} with the collective field $\theta(x)$ in the group-element basis.\footnote{%
More precisely, 
$\th_i \to \theta(x)$ with $x_i = \frac{i}{N}-\frac{1}{2} \to x$ in the large-$N$ limit.%
}
Consequently, Eq.~\eqref{eq:tilde_rho_saddle-pt} provides a direct saddle-point derivation of the relation in Eq.~\eqref{eq:Dutta-Gopakumar_rel}.

Now we move on to complete the computation of $\tilde\r(h)$, which will lead to the VKLS shape.
So far, we have focused on the interval $\abs{h}<2t$, where the two saddle points are complex conjugates and the kernel contains an oscillatory phase.
In this interval, the density satisfies $0<\tilde\rho(h)<1$.  
At the endpoints $h=-2t$ and $h=2t$, the two saddle points collide on the real axis and
the density $\tilde\rho(h)$ becomes $1$ and $0$, respectively.

\begin{figure}[t]
    \centering
    \includegraphics[width=0.75\textwidth]{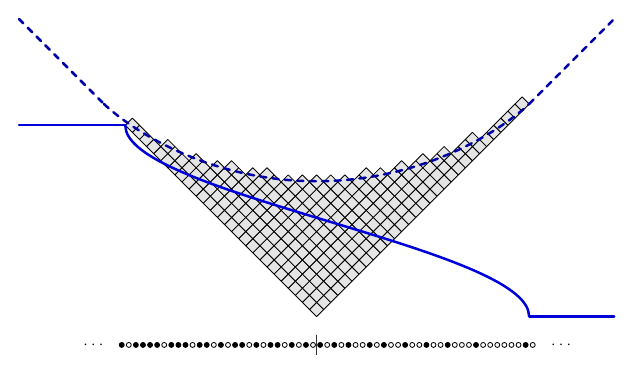}
    \caption{Schematic picture of a Young diagram and a Maya diagram in the scaling limit $L\to\infty$. The dotted and solid curves represent the VKLS limit shape $\Omega(h)$ and the particle density $\tilde \rho(h)$, respectively.}
    \label{fig:VKLS}
\end{figure}

For $h<-2t$ or $h>2t$, the saddle points \eqref{eq:saddle_points_kernel} become real.
We can show that
\begin{align}
\tilde\rho(h)=
\begin{cases}
0 & \text{ for } h>2t,
\\
1 & \text{ for } h<-2t.
\end{cases}
\end{align}
The limiting Maya diagram is no longer partially occupied in these regions:
all fermion levels are occupied to the left of the support ($h<-2t$),
while no fermion levels are occupied to the right ($h>2t$).
See Fig.~\ref{fig:VKLS}.

The saddle-point computation leading to this result can be summarised as follows.
For $h>2t$, the saddle points are located on the positive real axis: $z_+ >1$ and $0<z_-<1$.
The steepest descent directions of the integrand for the $z$-integration at $z_+$
and the $w$-integration at $z_-$ are both parallel to the imaginary axis.
Therefore, there is no obstruction when deforming the original contours to the ``steepest descent contours'' in this case. 
The contributions from the saddle points are exponentially suppressed since the real parts of the exponents are
negative at the saddle points for both the $z$- and $w$- integrations.
For $h< -2t$, while the contributions from the saddle points are exponentially suppressed, there is a contribution 
from the pole arising from the deformation of the contours.
This contribution takes the form \eqref{eq:kernel_contribution_from_pole}, with $\mathcal{C}_h$ now coinciding with the unit circle. 
Setting $\d = 0$, this integral becomes $\frac{1}{2\pi \ii}\oint \frac{\dd w}{w} = 1$.

To summarise, the resulting density on the Maya diagram is 
\begin{equation}
    \tilde\rho(h)
    =
    \begin{cases}
    \displaystyle
    1
    \quad&
    h < -2t,
    \\
    \displaystyle
    \frac{1}{\pi}\arccos(\frac{h}{2t}) 
    \quad&
    -2t \le h \le 2t,
    \\
    \displaystyle
    0
    \quad&
    h > 2t.
    \end{cases}
\end{equation}
This result is identical to that derived in Sec.~\ref{sec:derivation1}, except for the shift 
of the variable $h\to h-1$. We see that the parameter $t$ corresponds to the parameter $\xi$ there.

As already discussed below \eqref{eq:ZGWW}, 
computation of $\tilde{\rho}(h)$ for the GWW model is equivalent to
computation of $\tilde{\rho}(h)$ for the Gaussian matrix model.
Let us substantiate this point by using the Schur measure.
Writing 
\begin{equation}
    \hat{n}_r = \psi_{-r}\psi^*_r,
    \qquad
    \hat{n}_r\ket{\lambda} = n_r(\lambda) \ket{\lambda},
\end{equation}
and the fermion density as 
\begin{equation}
    \tilde\rho^{(X,Y)}(r)
    = 
    \frac{1}{Z(X,Y)} \sum_\lambda n_r(\lambda) s_\lambda(X) s_\lambda(Y),
\end{equation}
the corresponding density operator $\tilde{\rho}_h$ 
in the Gaussian matrix model becomes,
\begin{align}
    \tilde\rho_{\rm GMM}(r) 
    &=
    \frac{1}{Z_{\rm GMM}}
    \sum_\lambda n_r(\lambda) W_\lambda
    \\
    &=
    \frac{1}{Z_{\rm GMM}} 
    \int \qty[\dd\Phi] \ee^{-S_{\rm HS}(\Phi)} \sum_\lambda n_r(\lambda) s_\lambda(\Phi)s_\lambda(\Phi^*)
    \\
    &=
    \frac{\displaystyle\int \qty[\dd\Phi]\,\ee^{-S_{\rm HS}(\Phi)} 
    Z(\Phi,\Phi^*)\tilde\rho^{(\Phi,\Phi^*)}(r)}{\displaystyle\int \qty[\dd\Phi]\, \ee^{-S_{\rm HS}(\Phi)} 
    Z(\Phi,\Phi^*)}.
\end{align}
where we explicitly wrote the Miwa variables for $\tilde\rho^{(\Phi,\Phi^*)}(r)$.
This density can be directly evaluated at large $N$ because the integration over $\Phi$ localises 
at its saddle point in the large $N$ limit: 
\begin{equation}
    \tilde\rho_{\rm GMM}(r)
    \overset{N\to\infty}{=}
    \tilde\rho^{(\tau,\tau^*)}(r).
\end{equation}
This implies that, in the large-$N$ limit, the fermion density of the Gaussian matrix model is obtained by that of the Schur measure combined with the self-consistency condition \eqref{eq:self_consistency_tn_vs_an_un}.

Correspondingly, the limit shape of the tilted Young diagrams coincides with the VKLS shape~\cite{LS,VK} (see Fig.~\ref{fig:VKLS}):
\begin{equation}
    \Omega(h)
    =
    \int_h^\infty \dd{h'}\ 
    \qty(1-2\tilde\rho(h'))
    =
    \begin{cases}
    \displaystyle
     \frac{2}{\pi}\qty(\sqrt{4t^2-h^2}+h\arcsin(\frac{h}{2t}))
     &
     \abs{h} \le 2t,
     \\
     \displaystyle
     \abs{h} 
     &
     \abs{h} > 2t.
    \end{cases}
\end{equation}

We also comment on the relation to the standard probabilistic derivation.
For the GWW model, the Schur measure \eqref{eq:Schur_measure} reduces
to the Poissonised Plancherel measure,
$
\mu_{\rm PP}^{(t)}(\lambda)
= \ee^{t^2} t^{2\abs{\lambda}}(\dim \lambda/\abs{\lambda}!)^2
$.
The same saddle-point analysis of this measure was performed in
Ref.~\cite{BorodinOkounkovOlshanski2000}.
A $q$-deformed extension of this correspondence was studied in Ref.~\cite{Dutta:2021zdp}.
The CUE with potential deformation, or equivalently the Schur measure with
Miwa variables $t$ and $\tildet$, gives a natural extension
of the GWW model.
For such deformations, multicritical behaviour and the corresponding
higher-order scaling limits have been studied in
Refs.~\cite{Kimura:2020sud,Kimura:2021lrc,BeteaBouttierWalsh:2024}.

\subsection{Algebraic approach to saddle-point solution}
\label{sec:algebraic_approach}

The Fock-space description provides a way to carry out the saddle-point analysis in the basis of $\U(N)$ representations.
One may ask how the Polyakov loops in various representations, $\ev{s_\lambda(U)}$, can be computed within this representation basis.
In this subsection, we demonstrate the power of the free fermion representation by showing that such a computation can indeed be performed and that it reproduces the known solutions for the so-called ungapped phase.

This computation also connects the free-fermion description with
the algebraic structure introduced in Sec.~\ref{sec:dominant_YD} to define the dominant representations.
There, the key object was the two-variable generating function $Z_\b(U,V)$.
Its character expansion,
\begin{equation*}
    Z_\b(U,V) = \sum_\mu\sum_\nu Z_{\b \mu\n}\, s_\m(U)s_\n(V^{-1}),
\end{equation*}
combined with basic representation-theoretic quantities, namely the Littlewood--Richardson coefficients,
provides a way to compute the Polyakov loops. 
We shall see that the method based on the Schur measure reproduces this same structure.

Proceeding as in Sec.~\ref{sec:Schur_measure}, we find
\begin{align}
    \ev{s_\lambda(U)}
    &=
    \frac{1}{Z(t,\tildet)}
    \int_{\mathrm{U}(N)}\dd U\, s_\lambda(U) \exp[\sum_n\qty(t_n\tr U^n + \tildet_n \tr U^{-n})]
    \\
    &=
    \frac{1}{Z(t,\tildet)} \int_{\mathrm{U}(N)}\dd U\, s_\lambda(U) 
    \qty[\sum_\mu s_\mu(X)s_\mu(U)]
    \qty[\sum_\nu s_\nu(Y)s_\nu(U^{-1})]
    \\
    &=
    \frac{1}{Z(X,Y)} \sum_{\mu,\nu} s_\mu(X) s_\nu(Y) \int \dd U s_\lambda(U) s_\mu(U) s_\nu(U^{-1})
    \\
    &=
    \frac{1}{Z(X,Y)} \sum_{\mu,\nu} c_{\lambda\mu}^\nu\, s_\mu(X) s_\nu(Y),
\end{align}
where $c_{\lambda\mu}^\nu$ are the Littlewood--Richardson fusion coefficients.
We have used the fusion rule \eqref{eq:fusion_schur} and the orthonormality condition of the Schur polynomials.
Using the Fock-space description of $s_\lambda(X)$ together with \eqref{eq:fusion_coeff_algebraic}, we can evaluate the sum by inserting 
$
    1 = \sum_\lambda \ketbra{\lambda},
$
which yields
\begin{align}
    \ev{s_\lambda(U)}
    =
    \frac{1}{Z(X,Y)} \sum_{\mu,\nu} c_{\lambda\mu}^\nu\, s_\mu(X) s_\nu(Y)
    =
    \frac{1}{Z(X,Y)} \sum_{\mu,\nu} 
    \braket{X}{\mu}\mel{\mu}{\mathcal{A}_\lambda}{\nu}\braket{\nu}{Y}
    =
    \frac{\mel{X}{\mathcal{A}_\lambda}{Y}}{\braket{X}{Y}}.
\end{align}
Thus, we have rewritten the Polyakov loops in various representations, originally expressed as 
an integral over the group manifold, in terms of the representation basis.
Note that the creation operator $\mathcal{A}^\dagger_\lambda$ we introduced in \eqref{eq:creation_op_boson} 
leads to a remarkably concise form in the bra-ket notation.

Similar results can be obtained for a wider class of observables.
Recall that any square-integrable class function $f(U)$ on $\U(N)$ can be expanded in
terms of the characters:
\begin{equation}
    f(U) = \sum_{\lambda} f_\lambda s_\lambda(U).
\end{equation}
The expectation value of such $f(U)$ is given by 
\begin{equation}
    \ev{f(U)} 
    =
    \frac{1}{Z(X,Y)} \sum_{\l, \mu,\nu} f_\lambda c_{\lambda\mu}^\nu\, s_\mu(X) s_\nu(Y).
\end{equation}
For the case of the multiply-wound Polyakov loops, the Frobenius formula \eqref{eq:Frobenius_formula} yields
\begin{equation}
    \ev{\tr U^n} 
    = 
    \sum_{\mu\in \Rep{S_n}} \chi_\mu(C_\sigma) \ev{s_\mu(U)}
    =
    \sum_\mu \chi_\mu(C_\sigma) \frac{\mel{X}{\mathcal{A}_\mu}{Y}}{\braket{X}{Y}},
\end{equation}
where $\sigma$ is the conjugacy class of $S_n$ consisting of a single $n$-cycle, \textit{i.e.}, the partition with $\sigma_1 = n,~\sigma_{j\ge2}=0$.
We emphasise that $\mu$ in the sum runs over
the representations of the permutation group $S_n$ rather than $\U(N)$.
In the last expression, the bosonic mode operators and characters of the permutation group factorise out from the matrix element, and one can eliminate the characters using the orthogonality relation \eqref{eq:orthogonal_condition_character_perm}.
The commutation relations among the vertex operators and the bosonic mode operators \eqref{eq:comm_vertex_bos-mode} yield 
\begin{equation}
    \ev{\tr U^n} 
    = 
    \frac{\mel{\emptyset}{\Gamma_+(X)\rho_{n}\Gamma_-(Y)}{\emptyset}}{\braket{X}{Y}}
    =
    \tr Y^n
    =
    n \tildet_n.
\end{equation}
Similarly, we also have
\begin{equation}
    \ev{\tr U^{-n}}
    =
    \tr X^n
    =
    n t_n ,
    \label{eq:trU^n_algebraic}
\end{equation}
and therefore,
\begin{equation}
    \rho(\theta) 
    =
    \frac{1}{2\pi}
    \qty[1+\sum_{n=1}^\infty n\qty(t_n \ee^{-\ii n \theta} + \tildet_n \ee^{\ii n \theta})],
    \label{eq:rho(theta)_algebraic}
\end{equation}
which reproduces the so-called ungapped solution of the unitary matrix model \eqref{eq:CUE_pot-deformation}. 
(See, for example, Sec.~8.3 of Ref.~\cite{Marino:2015yie}.)

For the GWW model with $t_1 = \tildet_1 = t$ and $t_{n\ge2} = \tildet_{n\ge2} = 0$, 
the partially-deconfined phase of the original model is also correctly reproduced in this computation: 
Combining the self-consistency condition~\eqref{eq:self_consistency_tn_vs_an_un} with the fact that $a_1 = 1$ at $T = T_{\rm H} = T_{\rm GWW}$, one can directly identify the parameter $t$ with the Polyakov loop $u_1$.
Since the ungapped phase of the GWW model is realised for $0 \le t \le \frac{1}{2}$, the eigenphase distribution~\eqref{eq:rho(theta)_algebraic} coincides with that of the original model~\eqref{eq:rho(theta)_partially_deconfined}.

\subsection{Comment on the GWW transition}
\label{sec:large-N-transition-comment}
In the previous section, we considered the solution where the eigenphase distribution is non-zero everywhere on $[-\pi, \pi]$, \textit{i.e.}, the ungapped solution.
For the GWW model, the well-known GWW transition at $t=\frac{1}{2}$ causes a gap to open in the distribution.
Here, we discuss the gapped phase above the GWW transition
(or, equivalently, the deconfined phase $T>T_{\rm GWW}$ of the Gaussian matrix model)
and examine how the GWW transition may manifest itself
in the representation basis and affect the validity of the present algebraic approach.

This transition is associated with the trace operators $\tr U^n$ condensing to develop expectation values of order $N$ for $n \geq 2$.
In the representation basis, the dominant Young diagrams become large enough so that the number of rows reaches $N$, the maximum number of rows allowed for a $\U(N)$ representation.\footnote{%
For $\SU(N)$ gauge group, the maximum value is $N-1$. 
The difference is irrelevant in the large-$N$ limit.%
}
The resulting shape of the dominant Young diagrams has a characteristic ``capped'' feature: 
qualitatively it looks as if the lower part of some smooth shape
is abruptly cut off~\cite{Douglas:1993iia, Dutta:2007ws};
see Fig.~\ref{fig:schematic_YD_partially_and_completely_deconfined}.

In Ref.~\cite{Dutta:2007ws}, both $\r(\th)$ and $\tilde\r(h)$ were computed 
for the deconfined phase of the simplified model discussed in Sec.~\ref{sec:derivation1}.\footnote{%
Note that both $\r(\th)$ and $\tilde\r(h)$ for the simplified model differ from those
for the original Gaussian matrix model computed in Refs.~\cite{Sundborg:1999ue, Aharony:1999ti}.%
}
The result for $\tilde\r(h)$ is 
\begin{equation}
    \tilde\r(h) 
    =
    \frac{1}{\pi} \arccos\qty(\frac{h-1}{2\xi} + \frac{(\xi-1/2)^2}{2\xi h}),
    \label{eq:tilde_rho_cappedYD}
\end{equation}
for $h\in[q,p]$ with $\sqrt{q}=\sqrt{\xi}-\frac{1}{\sqrt{2}}$ and $\sqrt{p} = \sqrt{\xi} + \frac{1}{\sqrt{2}}$.
This density indeed corresponds to the capped shape of the dominant Young diagrams.
Interestingly, it was observed that the relation \eqref{eq:Dutta-Gopakumar_rel} holds 
between $\tilde\r(h)$ and $\r(\th)$ even in this deconfined phase, where $\r(\th)$ is gapped.

We have not succeeded in reproducing the saddle-point solution of the Gaussian matrix model 
from the present algebraic approach.
In fact, in its current form, the free-fermion formulation cannot produce the capped profile;
it is an intrinsically finite-$N$ effect, arising from the constraint that
the number of rows cannot exceed $N$.
In terms of the Maya diagram, the bound on the number of rows means that one cannot place a white circle (a ``hole'') below a minimum level $r\approx -N/2$, whereas no such restriction exists for black circles.
One crucial source of this limitation is 
the range of representations summed over in the Frobenius formula.
In various steps of the analysis in this section,
the orthogonality condition of the symmetric-group characters is used as if all partitions of $\abs{\sigma} = n$ contributed.
This is harmless when $n$ is kept fixed in the large-$N$ limit.
However, near and beyond the GWW transition, the relevant windings and Young diagrams scale with $N$; consequently, the set of representations of $S_n$ appearing in the Frobenius formula is no longer compatible with the set of allowed $\U(N)$ representations.~\footnote{%
A related discussion can be found in Ref.~\cite{Chattopadhyay:2019pkl}.%
}

Thus, while the partially deconfined phase can be analysed quantitatively 
using the strict large-$N$ limit,
the completely deconfined phase requires incorporating finite-$N$ effects.

This is reminiscent of a recent proposal which 
relates the GWW transition to trace relations~\cite{Hanada-Holden-Murthy-Sundborg}.\footnote{
The authors would like to thank Masanori Hanada for introducing us to this interesting idea through a presentation (\url{https://indico.yukawa.kyoto-u.ac.jp/event/77/contributions/1759/}) and subsequent private discussions, and for also sharing the draft of their work.
}
In the operator formalism of the Gaussian matrix model,
long-string states, or equivalently long traces in the matrix-model language, 
become relevant as temperature increases.
(See the explanation at the end of Sec.~\ref{sec:thermodynamics_GMM}.)
For large but finite $N$, sufficiently long single-trace operators, such as \eqref{eq:singletrace}, are
not independent: they can be expressed in terms of multi-trace operators built from shorter traces through
trace relations.
The authors of \cite{Hanada-Holden-Murthy-Sundborg} proposed that these trace relations 
play a key role in the GWW transition, 
where single trace operators of length $L\sim N^2$ become relevant.

An important direction for future work is to clarify the 
relation between these two finite-$N$ effects: trace relations and
the constraint on possible Young diagrams in our approach.
This might allow us to extend 
the algebraic approach to the GWW transition and the deconfined phase by
incorporating the finite-$N$ effect.~\footnote{%
For the reduced model, Ref.~\cite{deMelloKoch:2026jat} recently incorporated the finite-rank constraint explicitly by restricting Young diagrams to have at most $N$ rows. 
The crossover that occurs when the VKLS shape approaches this finite-rank wall was also analysed at finite $N$. The smoothing of the crossover was shown to exhibit Tracy--Widom scaling.%
}

\section{Summary and Prospect}
\label{sec:summary}
In this paper, we have considered the idea of the dominant representations and Young diagrams
of large-$N$ matrix models 
from the perspective of partial deconfinement.

\begin{figure}[t]
    \centering
    \includegraphics[width=0.65\linewidth]{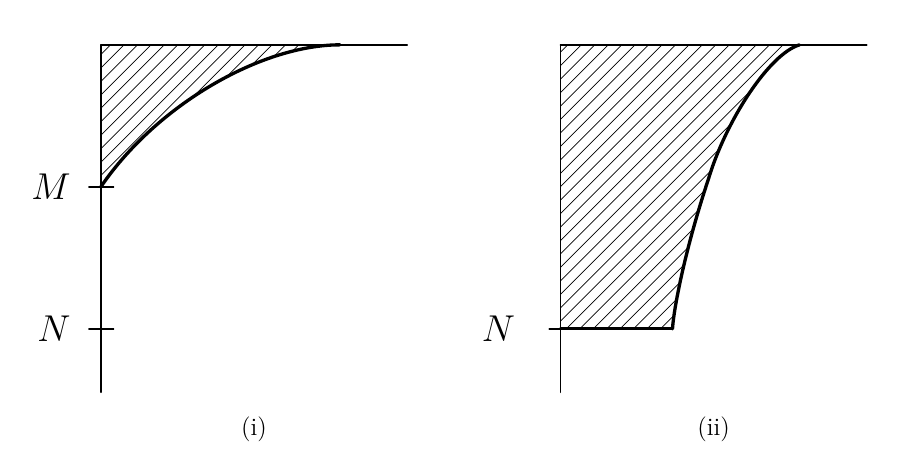}
    \caption{Schematic picture of the shape of the dominant Young diagrams. 
     In the confined phase, the dominant Young diagram is the empty diagram.
     (i) In the partially deconfined phase,
     the number of rows is $M$ (the size of the submatrix where deconfined degrees of freedom reside).
     (ii) In the deconfined phase, the shape has the capped profile.
    }
    \label{fig:schematic_YD_partially_and_completely_deconfined}
\end{figure}

The dominant Young diagrams offer a manifestly gauge-invariant way to measure
the size of the deconfined submatrix, $M$, which we expect to be applicable
to both free and interacting theories.
Fig.~\ref{fig:schematic_YD_partially_and_completely_deconfined}  schematically shows
the asymptotic shape of the dominant Young diagrams.
In the confined phase, the trivial representation dominates the thermal ensemble, 
and hence we have the empty Young diagram.
For the partially deconfined phase,
it is natural to identify the number of rows of the dominant Young diagrams with $M$,
reflecting the fact that the deconfined degrees of freedom reside in an $M\times M$ submatrix.
For the deconfined phase, 
the Young diagrams should have a capped profile where the height of columns
is saturated by $N$.
This picture is motivated by the fact that the number of rows of a Young diagram associated with
a $\U(N)$ representation is at most $N$.

We have proposed a prescription for defining the dominant Young diagrams for theories with interactions
in terms of a two-holonomy generating functional 
$Z_\b(U, V)=\sum_{\mu, \nu} Z_{\mu\nu} s_\mu(U) s_\nu(V^{-1})$. 
The expectation value of the Polyakov loop in a given representation is given by a sum
over Young diagrams of an expression 
involving $Z_{\mu \nu}$ and the Littlewood--Richardson coefficients.
The dominant Young diagrams are those diagrams which dominate the sum.

It is natural to expect that the shape of the dominant Young diagrams 
is connected to another gauge-invariant observable which captures the phase transitions: the eigenphase distribution of the thermal holonomy, $\r(\th)$.

In a sense that will be made more precise below, one can think of them as
being related by a Fourier transformation.
The eigenphase distribution is a function defined on the group manifold, 
while the Young diagrams correspond to the irreducible representations.
They are related through the character expansion, \textit{i.e.}, 
the Fourier transform on the group manifold.
The shape and $\r(\th)$ provide  
large-$N$ collective descriptions in the representation basis and the group-element basis, respectively.

We have substantiated these ideas by analysing a solvable example, the Gaussian matrix model,
using a mapping of the model to free fermions in two dimensions.
The shape of the dominant Young diagrams is computed from the number density of fermions 
$\tilde\rho(h)$.
It can be understood as a fluctuation around the Fermi sea represented by Maya diagrams,
which are related to Young diagrams by a simple geometric rule.
It is this fermion density $\tilde\rho(h)$, 
rather than the shape of the Young diagrams itself, that is directly dual to $\rho(\theta)$.

We have analytically computed $\tilde\rho(h)$ of the Gaussian matrix model, 
without introducing any simplifying assumptions, in the partially deconfined phase. 
The result is the so-called VKLS shape with the number of rows $M$, 
precisely confirming the above expectation depicted in Fig.~\ref{fig:schematic_YD_partially_and_completely_deconfined}~(i).

The same computation also serves as a direct derivation of a previously observed relation \eqref{eq:Dutta-Gopakumar_rel} between $\tilde\r(h)$ and $\r(\th)$.
The derivation features the following identification: 
the spatial coordinate and the momentum (the level) of the fermion system  
correspond to $\th$ and $h$, respectively.

Although we believe there should be a direct connection between $\r(\th)$ and $\tilde\r(h)$
for more general large-$N$ theories including those with interactions,
it is not clear to us whether the simple relation \eqref{eq:Dutta-Gopakumar_rel} applies as is.
There have also been several developments regarding this issue.
The relation holds as is for CUE with a potential deformation involving $n=2$ terms \cite{Chattopadhyay:2017ckc} and for the multicritical Schur measure \cite{BeteaBouttierWalsh:2024}.
A series of works~\cite{Matsuura:2023ova,Matsuura:2024gdu,Matsuura:2025khv} developed holonomy matrix models defined on graphs and investigated their large-$N$ phase structures and duality properties.
Building on these results, a recent work~\cite{Matsuura:2026epb} by the same authors reformulated the model on a cycle graph as a random partition model governed by the Schur measure and showed that \eqref{eq:Dutta-Gopakumar_rel} continues to hold as is for the model.
\footnote{
The authors would like to thank So Matsuura and Kazutoshi Ohta for insightful discussions and for sharing their results prior to submission to the arXiv.
}
It would be interesting to study whether this structure persists for holonomy matrix models on more general graphs.

We hope that our direct derivation of the relation \eqref{eq:Dutta-Gopakumar_rel} helps to clarify this issue.
In our derivation, we have seen that this relation is a manifestation of the fact 
that the variables $h$ and $\th$ are canonical conjugates. 
It is suggestive that \eqref{eq:Dutta-Gopakumar_rel} takes the form of a functional inversion $\rho(\theta) = h/2\pi$, which is reminiscent of the Legendre transformation.

Another possible approach to this issue is provided by 
the Hubbard--Stratonovich transformation used in our analysis.
It is natural to expect that, for theories with interactions,
the transformation would lead not to the simple Gaussian weight,
but to a more general nonlinear weight. 
This mapping may provide a useful starting point to find the direct relation between $\r(\th)$ and $\tilde\r(h)$ in interacting theories.

In the present work, we have not been able to derive, from the fermionic description of representations, the GWW transition that distinguishes the partially and completely deconfined phases.
However, we believe that the only missing ingredient is how to incorporate the finite-$N$ constraint when summing over representations of the permutation group and $\SU(N)$. 
This constraint may correspond, in the group-element language, to the emergence of trace relations near the critical temperature.
Since this issue is essentially a mathematical problem, it may be possible to resolve it systematically by using the knowledge of symmetric polynomials and combinatorics.

We also expect that this sensitivity to the finite-$N$ nature is closely related to the recent discussion of fortuitous states in the context of holographic dualities in superstring or M-theory~\cite{Chang:2013fba,Choi:2023tiq,Chang:2024zqi,Chang:2025d1d5,Belin:2025abjm}.
In that context, fortuitous states are BPS states whose existence depends on the precise finite value of $N$ and which therefore have no obvious large-$N$ counterpart.
The exactly solvable single-matrix model of Ref.~\cite{Chen:2025sum}, where fortuitous states are directly tractable in a unitary matrix integral as $N$ is varied, is closely related to our problem at a conceptual level.
Although the thermal large-$N$ transitions are not BPS-protected, both settings point to the same lesson: counting based on independent traces can differ from that in the finite-$N$ constrained Hilbert space once operators of length or charge of $O(N^2)$ become important~\cite{Hanada-Holden-Murthy-Sundborg}.

For now, we have employed the identification of an irreducible representation with a state in the Fock space of free fermions in two dimensions as a convenient mathematical tool for investigating the large-$N$ thermodynamics.
It would be interesting to ask whether the Fock-state description of irreducible representations is applicable to the emergent geometries in the context of gauge/gravity duality.
The interpretation of Young diagrams for the large-$N$ single Hermitian matrix model was already discussed in Ref.~\cite{Berenstein:2004kk}.
For the half-BPS sector of $\mathcal{N}=4$ super Yang--Mills theory in four dimensions, the relationship between Young diagrams and the information of extra holographic dimensions in the dual gravity description has been investigated.
In particular, Ref.~\cite{deMelloKoch:2008hen} argued that the Schur polynomials can be naturally translated into the boundary condition of LLM geometry~\cite{Lin:2004nb,Lin:2005nh} via a notion of free fermions.
Addressing these issues would help further understanding of the precise meaning or various aspects of the ``dominant representation.''

While strict phase transitions can only occur at large-$N$,
the picture of partial deconfinement may have some qualitative implications for 
and provide a useful framework for analysing finite-$N$ theories such as quantum chromodynamics (QCD).
Note that $N=3$ may be considered moderately large, since the expansion parameter is often $\frac1{N^2}$.
Of course, one must be careful about the difference between the large-$N$ limit and the infinite-volume limit from the viewpoint of their roles as thermodynamic limits.
Moreover, for QCD in our world, the confinement/deconfinement transition at small chemical potential is expected to be a crossover rather than a genuine phase transition (see, \textit{e.g.}, Ref.~\cite{Aoki:2006we}).
Nevertheless, phenomena reminiscent of partial deconfinement are found in the context of real-world QCD.
Numerical analyses of lattice QCD configurations~\cite{Hanada:2023krw} 
have revealed the following features:
in the low-temperature (``confined'') regime,
the Polyakov holonomy obeys the $\SU(3)$ Haar random distribution;
at higher temperatures, the Polyakov loops corresponding to
different representations of $\SU(3)$ begin to condense at different temperatures.
This is reminiscent of the existence of an intermediate phase characterised by
which degrees of freedom are excited in colour space---the defining feature of 
partial deconfinement.

More recently, several attempts have been made to understand aspects of the QCD phase diagram from a large-$N$ perspective (see \textit{e.g.} Refs.~\cite{McLerran:2007qj,Cohen:2023hbq}), 
including the Spaghetti of Quarks with Glueballs (SQGB) proposal~\cite{Fujimoto:2025sxx}.
The SQGB phase provides another intriguing analogy to partial deconfinement:
It describes an intermediate regime between the hadronic phase and the quark-gluon plasma, in which different degrees of freedom exhibit qualitatively different confinement properties. 
Although the thermodynamic limit of QCD is fundamentally different from the large-$N$ limit underlying partial deconfinement, these similarities suggest that the picture of partial deconfinement may provide a useful perspective for understanding the intermediate region of the QCD phase diagram~\cite{Hanada:2025rca}.
Note that, at finite density, one may encounter a richer phase structure than at zero or small chemical potential, as suggested by previous large-$N$ saddle-point analyses~\cite{Hands:2010zp,Hollowood:2011ep,Hollowood:2012nr} and from an observation of instanton condensation~\cite{Holden:2026bsf}.

From this perspective, an interesting next step is to reformulate the intermediate regime of QCD from the viewpoint of the representation basis developed in the present work. 
There are two conceptually distinct uses of representations that should be distinguished. 
One is the dominant Young diagrams, which characterise the thermal ensemble in the representation basis. 
The other is the representation $\lambda$ labelling the character $s_\lambda(U)$, whose expectation value probes deviations from the Haar-random distribution. 
In the Gaussian matrix model, these two descriptions are closely linked: for $T \le T_{\rm H}=T_{\rm GWW}$, only $u_1$ is excited, and its value fixes the scale of the dominant Young diagrams, or equivalently the size $M$ of the deconfined subsector. 
To study whether a similar link exists for QCD, a first step would be to study large-$N$
holonomy matrix models with single-trace terms, $\tr U^n$, in the action~\cite{Schnitzer:2004qt}, since QCD contains dynamical quarks, which are in the fundamental representation.
In such models, several Fourier modes $u_n$ can acquire non-zero expectation values as soon as the temperature becomes non-zero.
A natural first question is therefore how the information contained in the set $\{u_n\}$ is encoded in the dominant Young diagrams in these models. 
More fundamentally, one would like to identify the appropriate counterpart of the large-$N$ collective field $\tilde\r(h)$ for these models and study its relation to the eigenphase distribution.
Addressing these issues may provide some insight into partial deconfinement for more generic theories and the application of partial deconfinement to realistic QCD.

We hope that our work helps to establish the dominant Young diagrams as a useful tool for studying thermal gauge theories.

\acknowledgments
The authors thank Panos Betzios, Yiming Chen, Masanori Hanada, Yoshimasa Hidaka, Shota Komatsu, Stefano Kovacs, So Matsuura, Debangshu Mukherjee, Shinsuke Nishigaki, Kazutoshi Ohta, Ken Shiozaki, and Yuya Tanizaki for useful discussions and comments.
The authors are grateful to the organisers of the CERN workshop ``Matrix Quantum Mechanics for M-theory Revisited'', which helped initiate this work. 
H.S. is grateful to David Berenstein for a stimulating presentation based on the content of Ref.~\cite{Berenstein:2023srv} at the workshop.
H.W. is grateful to Taro Kimura, the author of \textit{Mathematical Physics of Random Matrices}, a textbook written in Japanese, from which he learned a great deal.
H.W. also thanks the organisers of the YITP conference ``Buenas Ideas on the QCD Phase Diagram'' (YITP-26-4) for providing an opportunity to present this work, and acknowledges financial support for travel expenses related to his participation.
The work of H.W. was partially supported by JSPS KAKENHI Grant Number 21J13014, 23K22489, and 24K00630.

\appendix

\section{Mathematical Preliminaries}

\subsection{Irreducible representations of \texorpdfstring{${\rm SU}(N)$}{SU(N)}}
\label{sec:irreps}

Irreducible representations of $\SU(N)$ are in one-to-one correspondence with 
Young diagrams (partitions) with at most $N-1$ rows.

A partition $\lambda$ is a non-increasing sequence of non-negative integers,
\begin{equation}
    \lambda 
    =
    \qty{\lambda_j\in \mathbb{Z}_{\ge 0}},
    \quad
    \lambda_1 \ge \lambda_2 \ge \cdots \ge 0.
    \label{eq:partition}
\end{equation}
We denote the size of the partition by $\abs{\lambda}$, 
\begin{equation}
    \abs{\lambda}
    =
    \sum_{j=1}^{\ell(\lambda)} \lambda_j ,
\end{equation}
where $\ell(\lambda)$ is the number of non-zero elements in $\lambda$.
A partition can be represented by a Young diagram. 
Note that $\ell(\lambda)$ is the number of rows and $\abs{\lambda}$ is the number of boxes of 
the Young diagram.

If there are two representations $\lambda$ and $\mu$, one can fuse them to obtain various representations $\nu$, as expressed by the fusion rule 
\begin{equation}
    \lambda \times \mu = \sum_\nu c_{\lambda\mu}^\nu \nu
    ,
    \label{eq:fusion_rule}
\end{equation}
where $c_{\lambda\mu}^\nu$ are the multiplicities of $\nu$ given by non-negative integers.
These coefficients are referred to as the fusion coefficients or the Littlewood--Richardson (LR) coefficients.

In this paper, we focus on large-$N$ matrix models. 
In particular, 
in Sec.~\ref{sec:derivation2}, 
we consider an alternative way of expressing representations in the strict large-$N$ limit~\cite{Jimbo:1983if,Okounkov2001InfiniteWedge}.
The difference between $\SU(N)$ and $\U(N)$ is usually negligible, particularly in the large-$N$ saddle-point analysis.

\subsection{Symmetric polynomials}
\label{sec:symm_poly}

The thermal holonomy $U \in \SU(N)$ plays a fundamental role in this paper.
We often consider its diagonalised form,
\begin{equation}
    U  \Leftrightarrow \diag(\ee^{\ii \theta_1},\dots,\ee^{\ii\theta_N}),
\end{equation}
characterised by the eigenphases $\theta_i \in [-\pi,\pi)$. 

The power-sum symmetric polynomial is defined as 
\begin{equation}
    p_n(U) = \tr U^n = \sum_{j=1}^N \ee^{\ii n \theta_j},
\end{equation}
which is the fundamental trace of group elements and is sometimes referred to as the multiply-wound Polyakov loop in the literature.
Since this is a class function, one can perform the so-called character expansion according to the celebrated Peter--Weyl theorem.
A key relation connecting the power sums to the characters is the Frobenius formula (see, \textit{e.g.}, Chapter~I, \S7, Eqs.~(7.2)--(7.5) of Ref.~\cite{Macdonald_sym_polynomial}):
\begin{equation}
    s_\lambda(U)
    =
    \sum_{\substack{\mu \\ {\rm s.t.}~ \abs{\mu}= \abs{\lambda}}}
    \frac{\chi_\lambda(C_\mu)}{z_\mu}
    p_\mu(U) , 
    \qquad
    p_\mu(U) 
    =
    \prod_{j=1}^{\ell(\mu)} p_{\mu_j}(U) , 
    \label{eq:Frobenius_formula}
\end{equation}
where 
\begin{equation}
    z_\lambda := \prod_j m_j!\, j^{m_j},
    \quad
    m_j = \#(\lambda_k = j \;|\; k \ge 1),
    \quad
    \lambda = (1^{m_1},2^{m_2},\dots).
\end{equation}
Here, $C_\mu$ is the conjugacy class of the permutation group $S_\abs{\mu}$ and $\chi_\lambda$ is the character of the permutation group.
The Schur polynomial is nothing but the character of $\U(N)$ in the representation $\lambda$; 
\begin{equation}
    s_\lambda(U) = \Tr_\lambda U.
\end{equation}
This can also be regarded as the Polyakov loop in the representation $\lambda$.

The characters satisfy orthogonality conditions:
For the characters of the permutation group, 
\begin{equation}
    \sum_{\substack{\mu \\ {\rm s.t.}~ \abs{\mu}= \abs{\lambda}}} 
    \frac1{z_\mu}
    \chi_\lambda(C_\mu) \chi_{\lambda'}(C_\mu) 
    = 
    \delta_{\lambda\lambda'}.
    \label{eq:orthogonal_condition_character_perm}
\end{equation}
For the Schur polynomials, 
\begin{equation}
    \int \dd U s_\lambda(U) s_{\lambda'}(U^\dagger) = \delta_{\lambda\lambda'}.
    \label{eq:orthogonal_condition_schur}
\end{equation}

The fusion rule for the Schur polynomials is given by  
\begin{equation}
    s_\lambda(X) s_\mu(X) 
    =
    \sum_\nu c_{\lambda\mu}^\nu s_\nu(X),
    \label{eq:fusion_schur}
\end{equation}
where the fusion coefficients are identical to those in \eqref{eq:fusion_rule}.

\section{Algebraic approach via doubling trick}
\label{sec:alg_approach_doubling}

We compute the weight $Z_{\mu\nu}(\beta)$ in \eqref{eq:partition_function_rep_generic} for the Gaussian matrix model.
As discussed, we lift the original $\SU(N)$ symmetry to $\SU(N)_{\rm L}\times \SU(N)_{\rm R}$ and consider the complex matrix model in the bifundamental representation.
Then, the partition function for fixed holonomies $U$ and $V$ can be decomposed as  
\begin{equation}
    Z_\beta[U,V] = Z_+[U,V] Z_-[U,V],
\end{equation}
where
\begin{equation}
    Z_{\pm}[U,V]
    =
    \exp[\sum_{n=1}^\infty \frac{\tilde a_n(\beta)}{n}\tr U^{\pm n}\tr V^{\mp n}],
\end{equation}
with 
$\tilde a_n(\beta) = \frac{D}{2}\,\ee^{-n\beta}$.
For simplicity, let us consider the $D=2$ case that corresponds to a model with a single complex field.
Note that a generalization to cases with $D > 2$ is straightforward.
Denoting $x = \ee^{-\beta}$, one can compute the first factor as 
\begin{equation}
    Z_+[U,V] 
    =
    \exp[\sum_{n=1}^\infty \frac{x^n}{n}\tr U^n \tr V^{-n}]
    =
    \sum_\lambda s_\lambda(xU) s_\lambda(V^\dagger)
    =
    \sum_\lambda x^\abs{\lambda} s_\lambda(U) s_\lambda(V^\dagger).
\end{equation}
In the middle, we have used that Schur polynomials are homogeneous of degree $\abs{\lambda}$.
The other factor $Z_-[U,V]$ is computed similarly.
On the other hand, the character expansion of $\SU(N)_{\rm L}\times \SU(N)_{\rm R}$ model yields 
\begin{equation}
    Z_\beta[U,V]
    =
    \sum_{\mu,\nu} Z_{\mu\nu}(\beta) s_\mu(U) s_\nu(V^\dagger),
\end{equation}
with a coefficient 
\begin{equation}
    Z_{\mu\nu}(\beta) 
    = 
    \int \dd U \dd V \,
    Z_\beta[U,V] s_\mu(U^\dagger) s_\nu(V)
\end{equation}
that depends on the details of the model.
Then, by using the fusion rule and orthogonality condition of Schur polynomials, one finds
\begin{equation}
    Z_{\mu\nu}(\beta) 
    =
    \sum_{\rho,\sigma}
    x^{\abs{\rho}+\abs{\sigma}} \, 
    c_{\sigma\mu}^\rho c_{\sigma\nu}^\rho.
    \label{eq:Z_RR'_gaussian_MM}
\end{equation}
We emphasise that non-zero components of the coefficients $c$'s are realised only 
when $\abs{\rho}=\abs{\sigma}+\abs{\m}$ and $\abs{\rho} = \abs{\s}+\abs{\n}$ 
due to the conservation law of the number of boxes.
Hence $Z_{\m\n}$ has a diagonal structure: it is non-zero only if $\abs{\m}=\abs{\n}$.
\footnote{
The coefficient $Z_{\mu\nu}(\beta)$ obtained here corresponds to the
Kronecker-coefficient data appearing in the $d$-matrix counting formulae of
Ref.~\cite{OConnor:2026zlf}.
}
Because of this diagonal nature, we obtain, using \eqref{eq:expectation_value_rep}, 
\begin{equation}
    \ev{s_\lambda(U)} = \sum_{\m, \n}c^\m{}_{\l\n} Z_{\m\n} = 0
\end{equation}
for any nontrivial representation $\lambda$,
since $c^\m{}_{\l\n}=0$ unless $\abs{\m}=\abs{\l}+\abs{\n}$.

This is not in conflict with the nonzero Polyakov loops found in the partially-deconfined and deconfined phases in Sec.~\ref{sec:thermodynamics_GMM}.
The vanishing of $\ev{s_\lambda(U)}$ for nontrivial $\lambda$ found here is an exact statement
which is valid for any finite value of $N$.
We have not taken into account the large-$N$ limit, hence the large-$N$ phase transitions
to the partially-deconfined and deconfined phases cannot be seen.
A different way of understanding this is the need for a microcanonical treatment of the 
partially-deconfined phase discussed in Sec.~\ref{sec:thermodynamics_GMM}. 
At the temperature $T_{\rm H}= T_{\rm GWW}$, there is a flat direction parametrised by
the Polyakov loop $u_1$. 
We must consider the states for a fixed value of $u_1$ (or equivalently energy) to specify 
a thermodynamic state of the partially-deconfined phase. 
This procedure is absent in the present analysis.

\begin{comment}
For the Gaussian matrix model, the coefficient~\eqref{eq:Z_RR'_gaussian_MM} can be directly computed employing the Frobenius formula, and the same calculation has been done to obtain \eqref{eq:partition_function_chisq}.
The result is 
\begin{equation}
    Z_{\mu\nu}(\beta)
    =
    x^{|\mu|}
    \sum_{\substack{\lambda \\ {\rm s.t.}~ \abs{\lambda}= \abs{\mu}}}
    \frac{D^{c(\lambda)}}{z_\lambda}
    \chi_\mu(C_\lambda)\chi_\nu(C_\lambda),
    \label{eq:Z_RR'_gaussian_MM_from_chisq}
\end{equation}
where $c(\lambda) = \sum_j m_j$ with $\lambda = (1^{m_1},2^{m_2},\cdots)$.
We see that \eqref{eq:partition_function_chisq} and \eqref{eq:Z_RR'_gaussian_MM_from_chisq} should match.
Presumably, one can prove this directly by using the Schur-Weyl duality which connects 
the $\U(N)$-symmetry and the permutation symmetries in a profound manner.
\end{comment}

\bibliographystyle{jhep}
\bibliography{Ref}

\end{document}